\documentclass[preprint,11pt]{elsarticle}

\usepackage[utf8]{inputenc}
\usepackage{graphicx}
\usepackage{amsmath}
\usepackage[version=4]{mhchem}
\usepackage{siunitx}
\usepackage{bm}
\usepackage{longtable,tabularx}
\usepackage{amssymb}
\usepackage{mathtools}
\usepackage{color}
\usepackage{xspace}
\usepackage{upgreek}
\usepackage{rotating}

\usepackage{graphicx}
\usepackage{dcolumn}
\usepackage{bm}
\usepackage{color}
\usepackage{upgreek}
\usepackage{multirow}
\usepackage{lipsum}
\usepackage{float}
\usepackage{ textcomp}
\usepackage[position=t,justification=justified,singlelinecheck=off]{subfig}
\usepackage{caption}
\usepackage[justification=justified]{caption}
\usepackage[colorinlistoftodos, color=blue!20!white,bordercolor=gray,textsize=tiny,textwidth=0.8in]{todonotes}
\usepackage{fancyhdr}
\usepackage{mleftright}
\usepackage[export]{adjustbox}
\mleftright

\def\bm#1{\mbox{\boldmath{$#1$}}}

\newsavebox{\astrutbox}
\sbox{\astrutbox}{\rule[-5pt]{0pt}{20pt}}

\journal{Journal of Aerosol Science}
\begin{document}

\begin{frontmatter}

\title{Multi-fidelity uncertainty quantification of particle deposition in turbulent pipe flow}


\author[label1,label3]{Yuan Yao}
\ead{yyao9@dow.com}
\author[label2]{Xun Huan}
\author[label1,label2]{Jesse Capecelatro}
\address[label1]{Department of Mechanical Engineering, University of Michigan, Ann Arbor, MI, 48109}
\address[label2]{Department of Aerospace Engineering, University of Michigan, Ann Arbor, MI, 48109}
\address[label3]{Engineering and Process Science, Core R\&D, The Dow Chemical Company, Lake Jackson, TX 77566}%

\date{\today}

\begin{abstract}
Particle deposition in fully-developed turbulent pipe flow is quantified taking into account uncertainty in electric charge, van der Waals strength, and temperature effects. A framework is presented for obtaining variance-based sensitivity in multiphase flow systems via a multi-fidelity Monte Carlo approach that optimally manages model evaluations for a given computational budget. The approach combines a high-fidelity model based on direct numerical simulation and a lower-order model based on a one-dimensional Eulerian description of the two-phase flow. Significant speedup is obtained compared to classical Monte Carlo estimation. Deposition is found to be most sensitive to electrostatic interactions and exhibits largest uncertainty for mid-sized (i.e., moderate Stokes number) particles.

\end{abstract}


\end{frontmatter}

\section{Introduction}
Particle deposition in turbulent wall-bounded flows can be observed in many engineering and environmental applications. Some examples include dry powder inhalers~\citep{begat2004cohesive,yang2013analysis,yang2015numerical}, dust ingestion in gas turbine engines~\citep{batcho1987interpretation,dunn1996operation,bons2017simple,sacco2018dynamic}, fluidized bed reactors~\citep{mikami1998numerical,van2008numerical,mahecha2009advances,pan2016cfd}, and indoor air quality~\citep{lai2000modeling,lai2002particle}. Many of these applications involve Geldart C-type particles (dust and powders with diameters less than $20$ $\upmu$m) for which cohesive forces such as van der Waals attraction and electrostatics become important~\citep{geldart1973types}. 

\begin{figure}[h!]
  \centering
  \includegraphics[width=.85\textwidth]{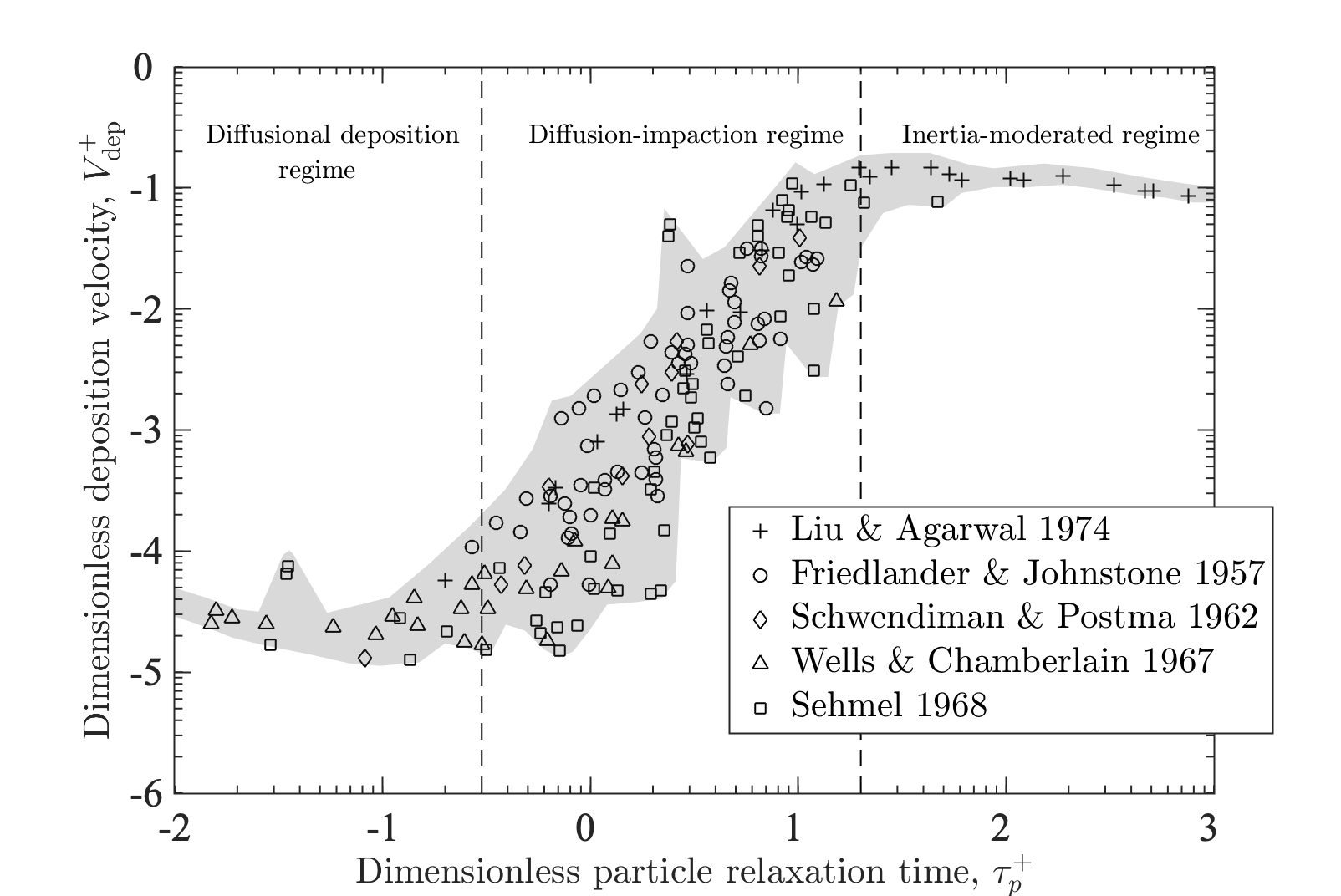}
  \caption{Compilation of experimental measurements of particle deposition in fully developed turbulent pipe flow. Shaded region highlights the 
  uncertainty from the variability in these measurement data.
  Figure adapted from~\citet{young1997theory}.}  
  \label{fig:dep_exp}
\end{figure}

Deposition is often measured in terms of the deposition velocity, $V_{\rm dep}$, defined as the particle mass transfer rate to a surface normalized by the bulk density of the particle phase~\citep{liu1974experimental,friedlander1957deposition,schwendiman1962turbulent,wells1967transport,sehmel1968aerosol}. Previous experimental measurements have observed three distinct regimes for the deposition velocity as a function of particle relaxation time, $\tau_p$ (see Fig.~\ref{fig:dep_exp} and definitions in \S~\ref{sec:dep_config}). For sub-micron particles, Brownian motion controls the deposition rate via turbulent diffusion, referred to as the turbulent diffusion regime. As the particle size increases, the deposition rate increases dramatically due to the interaction between inertial particles and fluid turbulent eddies. These particles are ejected from regions of high vorticity to the wall at a relatively high velocity, referred to as the diffusion-impaction regime. The third regime is known as the inertia-moderated regime where ballistic particles acquire sufficient momentum from turbulent eddies resulting in wall impact. In this regime, increased inertia results in a delayed response to the background turbulence and consequently the deposition rate reduces with increasing particle size.

Accurate predictions of the deposition process is challenging owing to its complex and multi-scale nature. The `free-flight’ or `stop-distance’ model proposed by~\citet{friedlander1957deposition} represents one of the first attempts at predicting the deposition velocity, which assumes particles directly deposit on the surface once they reach a `stop-distance' from the surface. However, it requires the particle velocity to be on the same order of the friction velocity at the stop distance, which often results in a significant over-prediction~\citep{young1997theory}. The model has been improved by many others~\citep{davies1966deposition,beal1970deposition,wood1981simple,papavergos1984particle}, but they typically require tuning of the so-called free-flight velocity and are only applicable to the diffusion-impaction regime. More recently, Lagrangian particle tracking has been adopted for deposition predictions. Velocity fluctuations of the carrier-phase turbulence can be obtained by numerically solving the Navier--Stokes equations~\citep{ounis1993brownian,brooke1994free,wang1996large}, estimated by empirical correlations~\citep{kallio1989numerical}, or using random walk models~\citep{gosman1983aspects,dehbi2008turbulent,gnanaselvam2021turbulent}. These approaches trade off between computational cost and accuracy. Alternatively, an Eulerian description of the particle phase provides an efficient means of predicting deposition. \citet{johansen1991deposition} found good agreement with experiments by combining a fully Eulerian approach with empirical closure models for the particle turbulence terms. More recently, \citet{guha1997unified,guha2008transport} proposed a one-dimensional Eulerian model derived from conservation laws that was shown to provide reasonably good predictions across all three regimes.


These models are all based on the assumption that particle cohesion is negligible compared to turbulent dispersion. However, cohesion is known to play an important role on transport of small particles in the turbulent diffusion regime (Geldart-C particles with diameters less than $20\,\upmu$m). \citet{guha2008transport} showed that the deposition rate in the first two regimes can be amplified by up to two orders of magnitude as a result of mirror charging at the wall alone using the aforementioned one-dimensional Eulerian model, whereas little effect is observed in the inertia-moderated regime. While incredibly valuable, the model only accounts for particle-wall electrostatic interactions and the particle distribution is assumed to be only a function of wall distance. In many systems, the spatial distribution of the particle phase can be significantly modified by electrostatic interactions~\citep{di2018aerodynamic,di2019mitigation,karnik2012mitigation}. In addition, particles are known to accumulate along low-speed streaks in turbulent boundary layers, defined as regions of lower-than-mean streamwise velocity~\citep{van1998direct,zonta2013particle,marchioli2003direct}. Such spatial inhomogeneities can alter particle near-wall deposition and are yet to be fully understood.

{\setlength{\tabcolsep}{9pt}
\begin{sidewaystable}[htbp]
 \begin{center}
  \begin{tabular}{llllll}
  \hline
  \hline
        Particle types & $d_p$ [$\upmu$m] & \multicolumn{2}{c}{Hamaker constant [$10^{\text{-}20}\,J$]} & Electric charge [$C$] & CoR\\ 
        & & $A_{11}$ & $A_{12}$ & &\\
       \hline
        SiO$_2$ (Silica)& $0.1 - 100$  & 6.5 & $5.37 - 13.7$ &  $8.0\rm{e}{\text{-}18} - 4.0\rm{e}{\text{-}16}$ & $0.62 - 0.77$\\
        SiO$_2$ (Quartz)& $ 0.5 - 15$ & $8.86$ & $7.59 - 12.1$ &  $8.0\rm{e}{\text{-}18} - 4.0\rm{e}{\text{-}16}$ & $ 0.19 $\\
        Al$_2$O$_3$ (Alumina)& $0.2 - 34$ & 15.2 & $7.90 - 21.1$ & $1.4\rm{e}{\text{-}18} - 2.4\rm{e}{\text{-}16}$ & $ 0.62 $\\
        CaSO$_4$ (Gypsum)& $ 0.7 - 20$ & $ - $ & $ - $ & $ - $ & $ 0.10$ \\	
        Dolomite& $ 2 - 40$ & $ 7.34 - 13.75$ & $ - $ & $ - $ & $ 0.15 $ \\	
        NaCl (Salt)& $ - $ & $ 6.48 $ & $ 6.45 - 10.3$ & $ - $ & $ 0.11 $ \\	
        Fe$_2$O$_3$ & $ - $ & $ 6.8 - 25$ & $ - $ & $ - $ & $ - $\\
        MgO & $ - $ & $ 10.6 - 12.1 $ & $ 8.84 - 14.2 $ & $ - $ & $ - $\\
        CaO & $ - $ & $ 12.4 $ & $ - $ & $ - $ & $ - $ \\
        TiO$_2$ & $ - $ & $15.3$ & $ 9.46 - 15.4$ & $ - $ & $ - $ \\
        \hline
        ARD & $1 - 40$ & $1.76 - 125$ & N/A & $1.08\rm{e}{\text{-}17} - 5.2\rm{e}{\text{-}16}$ & $ 0.34 - 0.76 $\\
        Fly ash & $1.5 - 80$ & $0.25 - 1.9$ & N/A & $8.7\rm{e}{\text{-}17} - 2.2\rm{e}{\text{-}16}$ & $ 0.16 - 0.75 $\\	
        Fine soil & $2 - 20$ & $1.05 - 513$ & N/A & $2.3\rm{e}{\text{-}17} - 2.5\rm{e}{\text{-}16}$ & $ 0.38 - 0.92 $\\	
       \hline
       \hline
  \end{tabular}
\caption{Size, Hamaker constant, charge, and coefficient of restitution (CoR) of common dust constituents reported in the literature~\cite{tsou1995silica,valmacco2016dispersion,kunkel1950static,bergstrom1997hamaker,exner2015powder,diaz2013adhesion,forsyth1998particle,faure2011hamaker,crosby2007effects,galbreath1999reducing,tabakoff1991effect,ontiveros2016effect,gilbert1991charge,crowe2019effects,singh2013predicting,li2015rebound,dong2013experimental,tsai1991elastic,bojdo2019simple,koper2020influence,goudarzy2016influence,tanaka2002study,lefevre2009calculation,yao2016method,resurreccion2011relationship,sandeep2021experimental,reagle2013measuring,moutinho2017investigation} of common natural particulates and dust constituents. Upper and lower bounds correspond to the  $10$-th and $90$-th percentile when distributions are given. ARD denotes Arizona road dust.}
   \label{table:dust}
 \end{center}
\end{sidewaystable}}

Significant uncertainty associated with cohesive force measurements and material properties of dust and powders introduce additional challenges. For example, the measurement of the Hamaker constant, which dictates the strength of van der Waals attraction, can span five orders of magnitudes for typical dust~\citep{vowinckel2019settling}. It remains an ongoing debate in the literature how to parametrize and measure cohesive forces accurately~\citep{ho2002modelling,lick2004initiation,breuer2015modeling}. 
As summarized in Table~\ref{table:dust}, particle properties exhibit large variations even at standard atmospheric conditions. Temperature effects and surface roughness further amplify these uncertainties~\citep{guha2008transport}. Nevertheless, most existing numerical studies consider monodisperse particles with single-valued charge~\cite{lu2015charged,lu2010clustering} and Hamaker constant~\cite{ho2002modelling,hartley1985role}. In general, accurately quantifying the underlying uncertainty requires many expensive simulations. 
Efficient methods 
for uncertainty quantification (UQ) 
are thus needed
to describe the input-output 
variability
between particle properties and deposition rate in a tractable manner---a key focus of the present work.

A core task of UQ is to propagate uncertainty from the sources (inputs) through a system in order to characterize the resulting uncertainty of other quantities of interest (QoIs) (outputs).
%
With advances in computational capabilities and a growing need to produce confidence information for mission- and decision-critical model predictions, UQ 
is becoming widely used to accompany
fluid dynamic studies~\citep[e.g.,][]{oliver2011bayesian,turnquist2019multiuq,bravo2021uncertainty,nili2021prioritizing,klemmer2021implied}. 
The primary approach to UQ revolves around Monte Carlo (MC) sampling \citep[e.g.,][]{Robert2004}, which is multi-query in nature and generally requires many numerical simulations to obtain estimates with acceptable accuracy \citep{papadopoulos2001uncertainty, bijl2013uncertainty}. Recently, multi-fidelity Monte Carlo (MFMC) methods have been introduced to leverage  low-fidelity models to accelerate the uncertainty propagation by occasionally making use of a high-fidelity model \citep{peherstorfer2016optimal,peherstorfer2018survey,Eldred2017}. The salient idea is to optimize the work distribution among models of different fidelity such that the overall MC error is minimized for a given computational budget. Another advantage of the MFMC framework is that the mean quantity-based optimization admits an analytical solution that can be directly applied for variance and sensitivity estimates \citep{qian2018multifidelity}. The MFMC approach was recently applied to thermally irradiated particle-laden turbulent flows, and exhibited a speedup of up to $10^{3}$ compared to the classical MC method~\citep{jofre2020multifidelity}. 

Directly related to the task of uncertainty propagation is variance-based global sensitivity analysis (GSA) \citep{Sobol2001,Saltelli2008}, which aims to compute Sobol' indices that decompose the total variability of an output quantity into contributions from the uncertainty of each input. In addition to providing a quantitative measure and ranking on the importance of each input from an uncertainty point of view, GSA is also useful for dimension reduction by guiding the removal of low-impact  stochastic (uncertainty) degrees of freedom. While MC algorithms have been developed to estimate Sobol' indices \citep{Sobol1990,Jansen1999,Saltelli1999,Saltelli2002}, they remain expensive and generally do not make use of multi-fidelity models.

In the present study, Sobol' indices are computed for the particle deposition rate in a turbulent pipe flow using MFMC. Two deposition models of different fidelities, direct numerical simulations and the one-dimensional Eulerian model proposed by~\citet{guha2008transport}, are used to expedite the evaluations. The system configuration is described in \S~\ref{sec:physics}, followed by a comparison of deposition rates predicted by these two models. The multi-fidelity framework is introduced in \S~\ref{sec:mfmc} along with GSA
using Sobol' indices to quantify the relative importance of cohesive forces for three different particle sizes.

\section{Particle deposition in fully-developed pipe flow}\label{sec:physics}
In this section, details on the two-phase flow configuration are provided, followed by an analysis of the various forces contributing to deposition. Deposition statistics obtained from a high-fidelity model and lower fidelity model are compared, and used to motivate the UQ study in the following section. Two-phase flow statistics associated with this configuration can be found in our previous work \citep{yao2021accurate,lo2022assessment}.

\subsection{System configuration}\label{sec:dep_config}
\begin{figure}[h!]
\centering
\subfloat[]{
\includegraphics[width=0.58\textwidth]{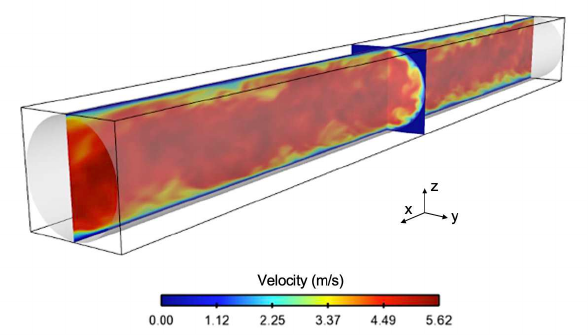}
\label{fig:vis-1}}
\subfloat[]{
\includegraphics[width=0.32\textwidth]{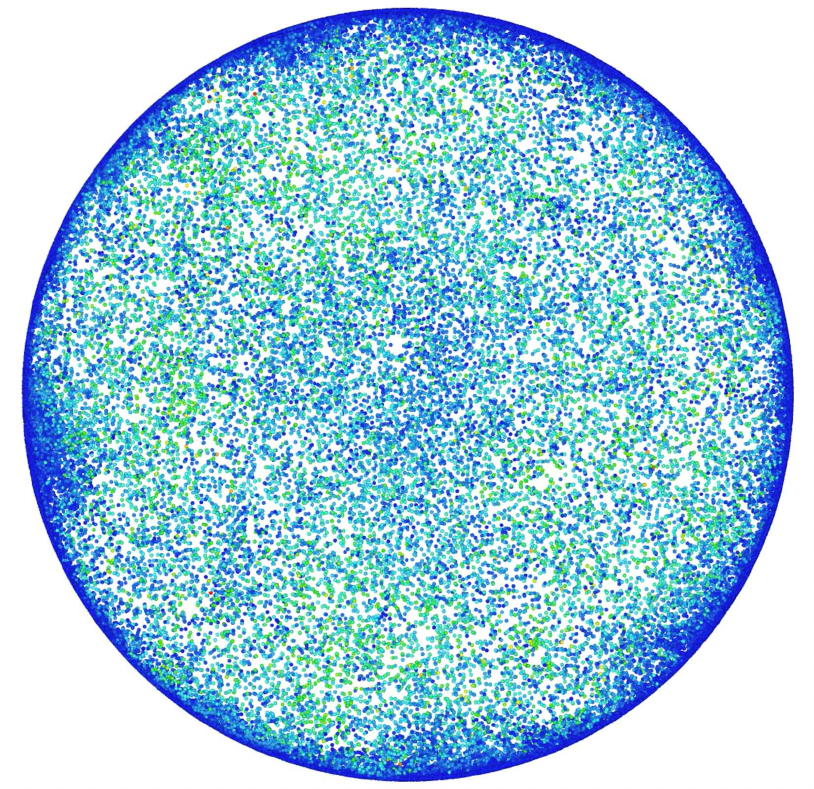}
\label{fig:vis-2}}
\caption{Direct numerical simulation of the turbulent pipe flow. (a) The pipe wall is depicted in gray. Color represents the instantaneous fluid velocity magnitude at a statistically stationary state. (b) Particle distribution colored by radial velocity (blue: $0$, red: $1\,$m/s) for $\tau_p^{+}=10$ at a statistically stationary state.}
\label{fig:config}
\end{figure}
The problem under consideration consists of a cylindrical pipe with diameter $D=20$ mm (see Fig.~\ref{fig:config}). The length of the pipe is $L=10D$ to ensure domain independent results. The pipe is periodic in the streamwise ($x$) direction. The fluid is taken to be air with kinetic viscosity $\nu=1.5\times 10^{-5}$ and bulk velocity $U_b = 4.0$~m/s, corresponding to a Reynolds number based on the friction velocity of $\rm{Re}_{\tau} = 180$ and a Reynolds number based on the bulk velocity $\rm{Re} = 5333$. Monodisperse spherical particles of three different diameters $d_p = 1.6$ $\upmu$m, $5$ $\upmu$m and $16$ $\upmu$m and density $\rho_p=2650$ kg/m$^3$ are considered. The simulations consist of $N=10^{6}$ particles to ensure a statistically significant sample size, corresponding to an average number density $n_p = 1.6\times 10^{10}$ m$^{-3}$. Particles are randomly distributed within the domain and reach a statistically stationary state where the particle mass flux at the wall is constant in the radial direction. Particle inertia is characterized by a non-dimensional particle response time $\tau_p^{+} = \tau_p / \tau_f$, where $\tau_p = \rho_pd_p^2/(18\rho \nu)$ is the particle response time and $\tau_f = \nu/u_*^2$ is the frictional time scale. The three particle sizes considered correspond to $\tau_p^{+} = 0.1, 1, {\rm and\ } 10$.

To investigate the effect of charge on particle dynamics and deposition, each particle is assigned a unique value $q_p$ that is proportional to the maximum possible charge $q_{\rm max}$ according to $q_p = \xi q_{\rm max}$, with {$0\le\xi\le0.1$}. 
The maximum charge is given by~\cite{howard1977fine}
\begin{equation}
q_{\max }=500 \times\left(1.6 \times 10^{-19}\right)\left(d_p / 10^{-6}\right)^{2}.
\end{equation}
In this study, we consider $\xi=0$ (no charge) and  $\xi=0.1$, representative of typical values for charge measured in dustry turbulent pipe flows~\citep{matsusaka2002electrostatic,matsusaka2006simultaneous,rodrigues2006measurement}. The Hamaker constant, $A$, is varied between $10^{-20}\,$J for weakly cohesive particles (e.g., silica) to $10^{-18}\,$J for strongly cohesive materials such as metal oxides~\citep{marshall2014adhesive}. Similarly, a non-dimensional parameter $\zeta = A/A_0$ is introduced with $A_0=10^{-18}\,$J. 
We consider {$0\leq\zeta\leq 1$.}

The deposition rate is typically characterized using the non-dimensional deposition velocity~\citep{kallio1989numerical,uijttewaal1996particle,matida2000statistical,marchioli2003direct,phares2006dns,zonta2013particle}, given by
\begin{equation}
V_{\rm dep}^{+} \equiv \frac{V_{\rm dep}}{u_*} = \frac{J_{w}}{\rho_{p m} u_*},
\end{equation}
where $J_w$ is the particle mass flux to the wall per unit area, $\rho_{pm}$ is the mean particle bulk density in the pipe and $u_* = \sqrt{\tau_w/\rho}$ is the friction velocity with $\tau_w$ the wall shear stress. Throughout the rest of this paper, $\zeta$ and $\xi$ are taken as input parameters
that introduce parametric uncertainty to the system, 
and $V_{\rm dep}^{+}$ is the system output or QoI. Thus, we will perform UQ and GSA targeting the QoI $V_{\rm dep}^{+}$.

\subsection{Relative importance of cohesive forces}
Before performing UQ and GSA,
we first analyze the order of magnitude in van der Waals and electrostatic forces in wall-bounded systems with different particle loadings.
The magnitude of the van der Waals force between two particles $i$ and $j$ of the same size is given by
\begin{equation}\label{eq:vdw}
F_{i j}^{\mathrm{vw}}(\delta_{ij})=\frac{A}{6} \frac{d_p^2\left(d_p+\delta_{ij}\right)}{\delta_{ij}^{2}(2d_p+\delta_{ij})^{2}}\left[\frac{\delta_{ij}\left(2d_p+\delta_{ij}\right)}{(d_p+\delta_{ij})^{2}}-1\right]^{2},
\end{equation}
where $\delta_{ij}$ is the distance between the particle surfaces, and $A$ is the Hamaker constant. Due to its short range nature, it is assumed that the force saturates at a minimum separation, $\delta_\text{min} = 1\times10^{-9}$ m and is cut off at $\delta_\text{max} = d_p/4$. The magnitude of the electrostatic force is given by Coulomb's law according to
\begin{equation}
\label{eq:coulomb}
F_{ij}^C(\delta_{ij}) = \frac{q_p^{(i)}q_p^{(j)}}{4\pi \epsilon_0 \delta_{ij}} =  \frac{q_p^{(i)}q_p^{(j)}}{4\pi \epsilon_0} \frac{1}{\big \vert \bm{x}_p^{(j)} - \bm{x}_p^{(i)} \big \vert^2},
\end{equation}
where $q_p^{(i)}$ and $q_p^{(j)}$ are the charges belonging to particles $i$ and $j$, respectively, and $\epsilon_0= 8.854 \times 10^{-12}$ $\text{F} \cdot \text{m}^{-1}$ is the vacuum permittivity. Details on the numerical discretization of these forces can be found in \ref{sec:hf_model}.

\begin{figure}[h!]
  \centering
  \subfloat[]{\includegraphics[width=.495\textwidth]{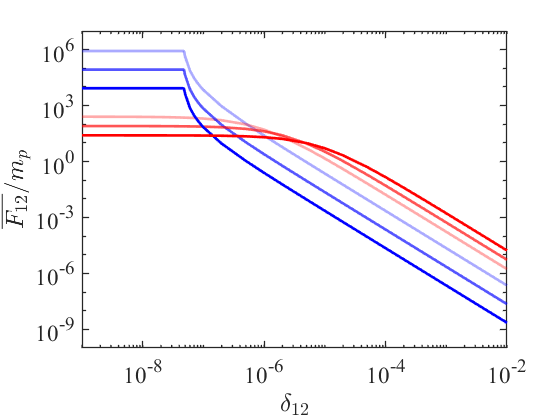}\label{fig:dep_analytical_a}}
    \subfloat[]{\includegraphics[width=.495\textwidth]{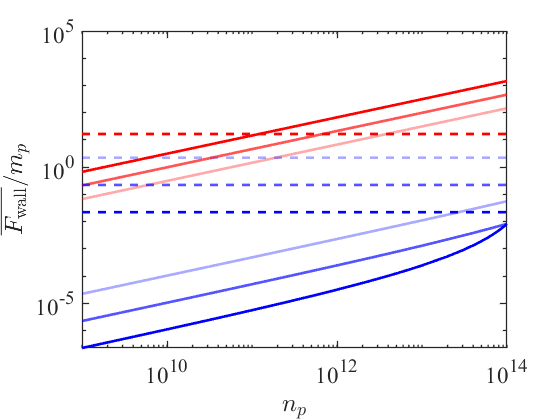}\label{fig:dep_analytical_b}}
  \caption{Particle acceleration due to electrostatics ($\color{red}{-}$) and van der Waals ($\color{blue}{-}$) as a function of (a) separation distance $\delta_{12}$ and (b) particle number density $n_p$. Larger particles are depicted as lines with higher color opacity. Particle-wall contributions are shown as dashed lines.}
  \label{fig:dep_analytical}
\end{figure}

Figure~\ref{fig:dep_analytical_a} shows the particle acceleration due to these two forces ($\xi = 0.1$ and $\zeta = 1$) as a function of particle separation distance via a direct evaluation of Eqs.~\eqref{eq:vdw} and~\eqref{eq:coulomb}. The van der Waals force is seen to be the dominant mode of deposition at short separation ($\delta_{12}<\mathcal{O}(10^{-6}\,{\rm m})$), but is surpassed by electrostatic forces at longer range. The influence of the van der Waals force decays with increasing particle size, whereas the electrostatic force results in higher particle acceleration for larger particles at long distances. 

From Fig.~\ref{fig:dep_analytical_a}, the average particle acceleration can be estimated for a given number density $n_p$. By assuming a random distribution of particles, the average distance between particles $\delta_{12} = n_p^{-1/3}$. Let $F_{12}(\delta_{12})$ denote the force between two particles given by Eqs.~\eqref{eq:vdw} and \eqref{eq:coulomb} as a function of separation distance. The domain-averaged particle-particle force is therefore $\overline{F}_{12} = F_{12}(1/{n_p}^{1/3})$. The particle-wall force $F_{\rm wall}(\delta_{12}) = 2F_{12}(\delta_{12})$ for van der Waals interactions and $F_{\rm wall}(\delta_{12}) = F_{12}(2\,\delta_{12})$ for electrostatic interactions, where image charging is assumed by placing an image particle of opposite charge mirrored at the same distance across the wall. The domain-averaged particle-wall force is estimated by integrating the force in the radial direction
\begin{equation}
\overline{F}_{\rm wall} = \int_0^{0.5D-r_{\rm cutoff}} F_{\rm wall}(0.5D-\delta_{12}) \frac{8\delta_{12}}{D^2} \ {\rm d}\delta_{12},
\end{equation}
where $r_{\rm cutoff} = \delta_{\min}$ for van der Waals and $r_{\rm cutoff} = d_p/2$ for electrostatics to avoid singularity. The accelerations due to estimated forces are compared in Fig.~\ref{fig:dep_analytical_b}. It can be seen that domain-averaged particle-particle contributions increase with $n_p$ while particle-wall contributions are insensitive to the particle loading. As a result, the relative importance of these four different forces strongly depends on $n_p$. For instance, the particle-particle electrostatic force surpasses particle-wall image charging when $n_p \gtrsim \mathcal{O}(10^{11})$. Due to the short-range nature of van der Waals interactions, however, the particle-wall contribution is always greater than the inter-particle van der Waals attraction. For the system considered herein ($n_p = 1.6\times 10^{10}$), the relative importance is ranked as particle-wall electrostatics, particle-particle electrostatics, particle-wall van der Waals, and particle-particle van der Waals, which is orders of magnitudes smaller than the former three.

\begin{figure}[h!]
  \centering
  \subfloat[]{\includegraphics[width=.495\textwidth]{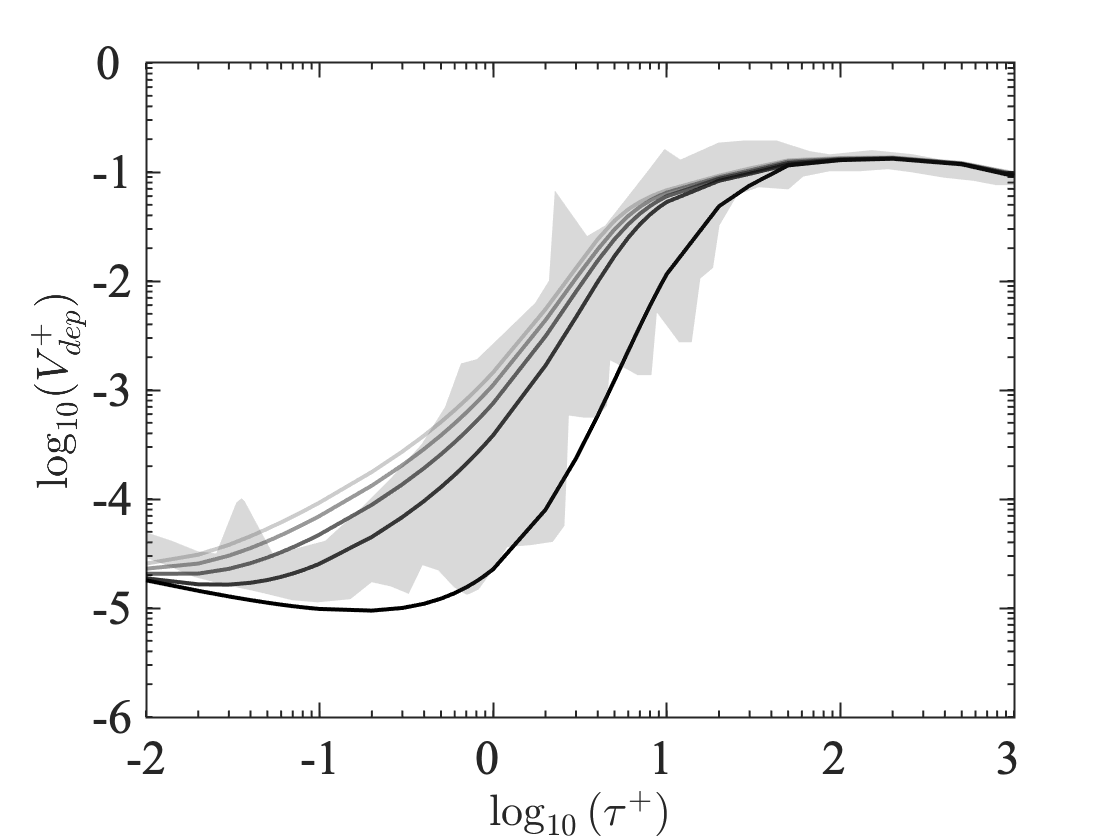}}
    \subfloat[]{\includegraphics[width=.495\textwidth]{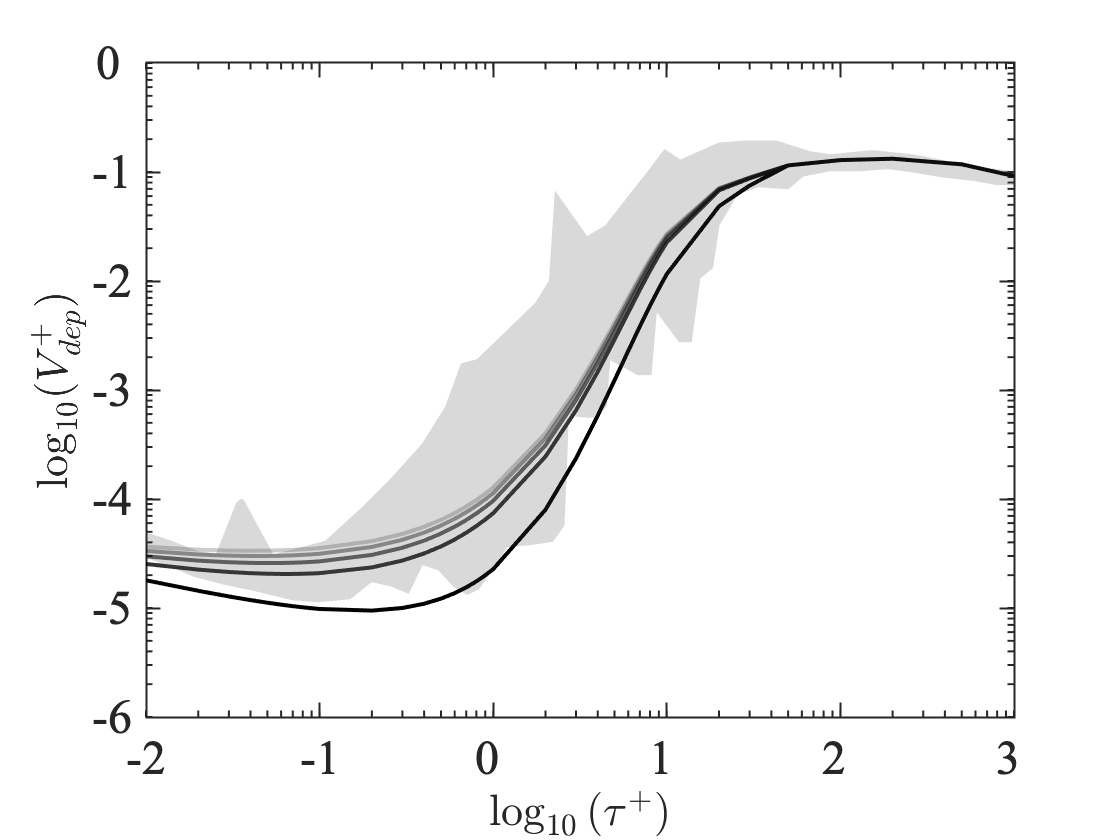}}
  \caption{Non-dimensional deposition velocity as a function of the non-dimensional particle response time predicted by the one-dimensional Eulerian model of~\citet{guha2008transport} for (a) electrostatics due to image charging with $\xi=0, 0.025, 0.05, 0.075, 0.1$ and (b) van der Waals attraction with $\zeta =0, 0.25, 0.5, 0.75, 1$. Higher values of $\xi$ and $\zeta$ are shown with increasing transparency. Shaded region same as Fig.~\ref{fig:dep_exp}.}
  \label{fig:dep_1d}
\end{figure}

The non-dimensional deposition velocity predicted by the one-dimensional Eulerian model of~\citet{guha2008transport} is shown in Fig~\ref{fig:dep_1d}. For the range of parameters considered, the electrostatic force is seen to have a larger effect on the deposition rate than the van der Waals force across all particles sizes, which is consistent with the direct evaluation of Eqs.~\eqref{eq:vdw} and \eqref{eq:coulomb} reported in Fig.~\ref{fig:dep_analytical_b}. In addition, the electrostatic force exhibits the largest enhancement in deposition for mid-sized particles (diffusion-impaction regime), for which the charge amount is significant and at the same time particles are most responsive to the turbulent eddies. Van der Waals interactions, on the other hand, primarily affect the deposition rate of small particles (turbulent diffusion regime).

\subsection{A comparison of high- and low-fidelity models}
\label{ss:comparison}

In this work, two model fidelities are considered to facilitate UQ and GSA
in the later sections. The high-fidelity model is taken to be a direct numerical simulation (DNS) of the fully-developed turbulent pipe flow presented in \S~\ref{sec:dep_config}. The one-dimensional Eulerian model proposed by~\citet{guha2008transport} is used for the low-fidelity model. Details of each model are described in \ref{sec:hf_model} and \ref{sec:lf_model}. In this section, the relative importance of cohesive forces on particle deposition is quantified, and a comparison is made between the high- and low-fidelity models. Multi-fidelity UQ is then presented in the following section.

\begin{figure}[h!]
\centering
\includegraphics[width=0.58\textwidth]{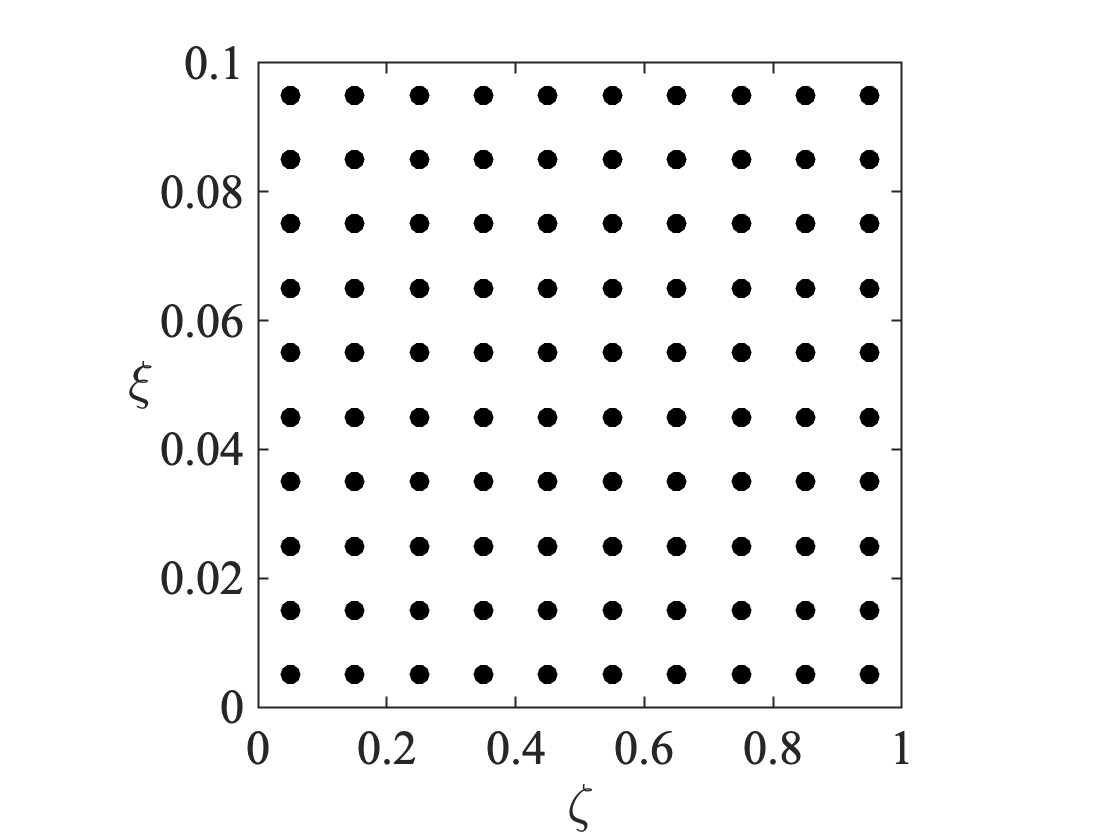}
\caption{Uniform probing of 100 pairs of input parameters with $\xi\in[0,\,0.1]$ and \textbf{$\zeta\in [0,\,1]$}.}
\label{fig:uq_inputs}
\end{figure}

The efficacy of MFMC for UQ of particle deposition depends on how strongly the two models correlate (as long as the low-fidelity prediction varies in a \emph{correlated} manner to its high-fidelity counterpart---not necessarily needing to be 
accurate---then it could be made into a good predictor to the high-fidelity behavior). 
To illustrate the correlation 
between the models 
and also to gain an initial physical intuition for the problem, in this section we evaluate a set of  high-fidelity (DNS) and low-fidelity (Eulerian one-dimensional model) simulations
systematically in the parameter space ($\xi$ and $\zeta$).
In particular, we endow uniform uncertainty distributions for both $\xi$ and $\zeta$ across their allowable ranges ($\xi\in[0,0.1]$,
$\zeta\in[0,1]$), 
representing a scenario where all values within their ranges are equally likely to occur. The uniform distribution appeals to the principle of maximum entropy \citep{Jaynes1957} that makes the fewest assumption in distribution formation
when upper and lower bounds are given. 
100 simulations are then performed for both models by uniformly probing the parameter space shown in Fig.~\ref{fig:uq_inputs}. 
We emphasize that this is not part of the MFMC or MC sampling, but rather a 
``brute-force'' 
illustration
using a tensorized sweep of the entire parameter space, which is  achievable here because the parameter space is only two-dimensional.

\begin{figure}[h!]
  \centering
  \subfloat[]{\includegraphics[width=.336\textwidth,valign=b]{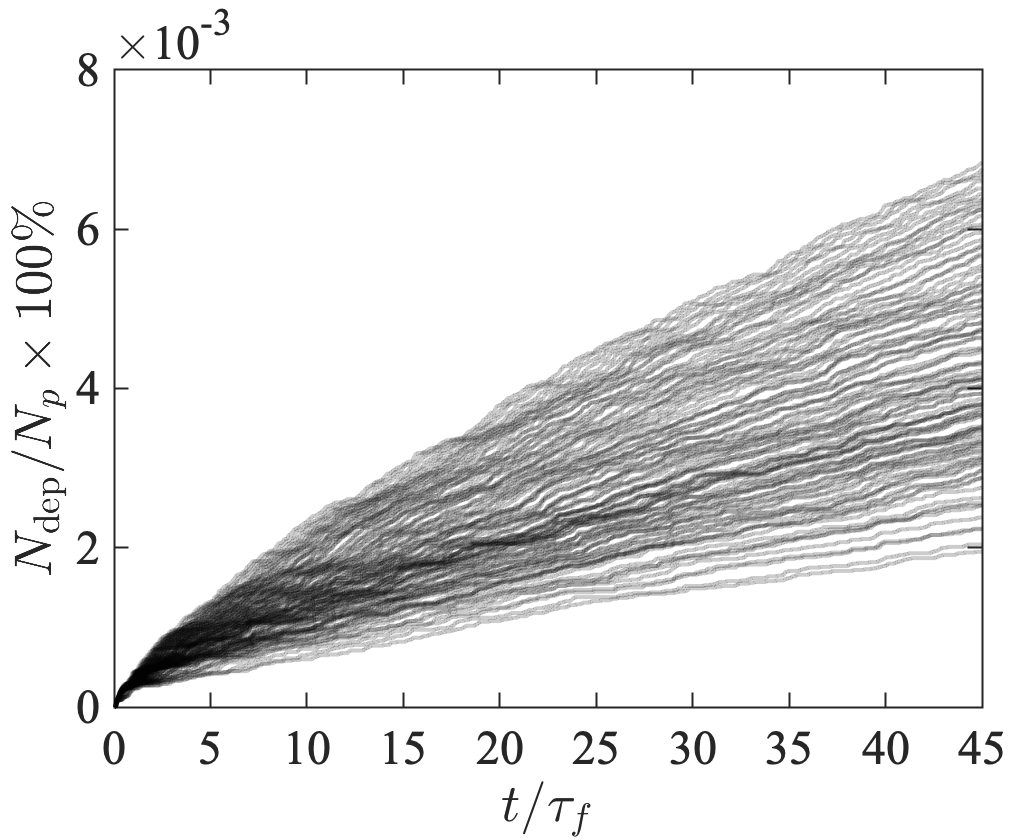}\label{dep-a}}
   \subfloat[]{\includegraphics[width=.332\textwidth,valign=b]{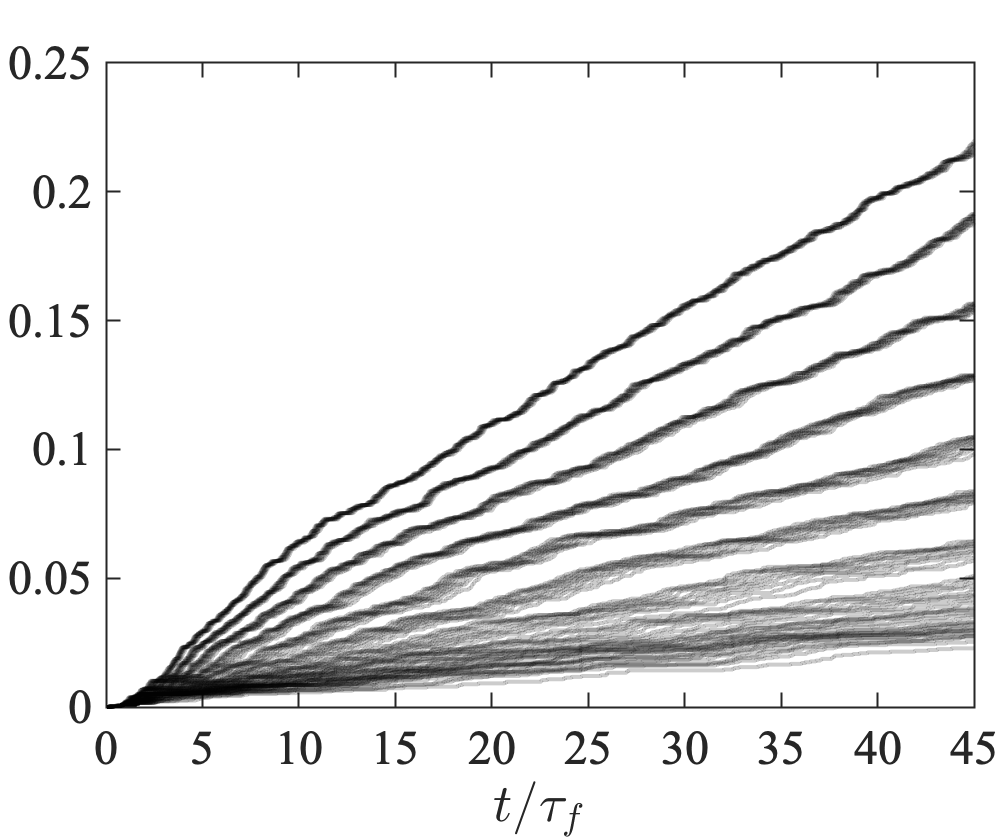}\label{dep-b}}
    \subfloat[]{\includegraphics[width=.328\textwidth,valign=b]{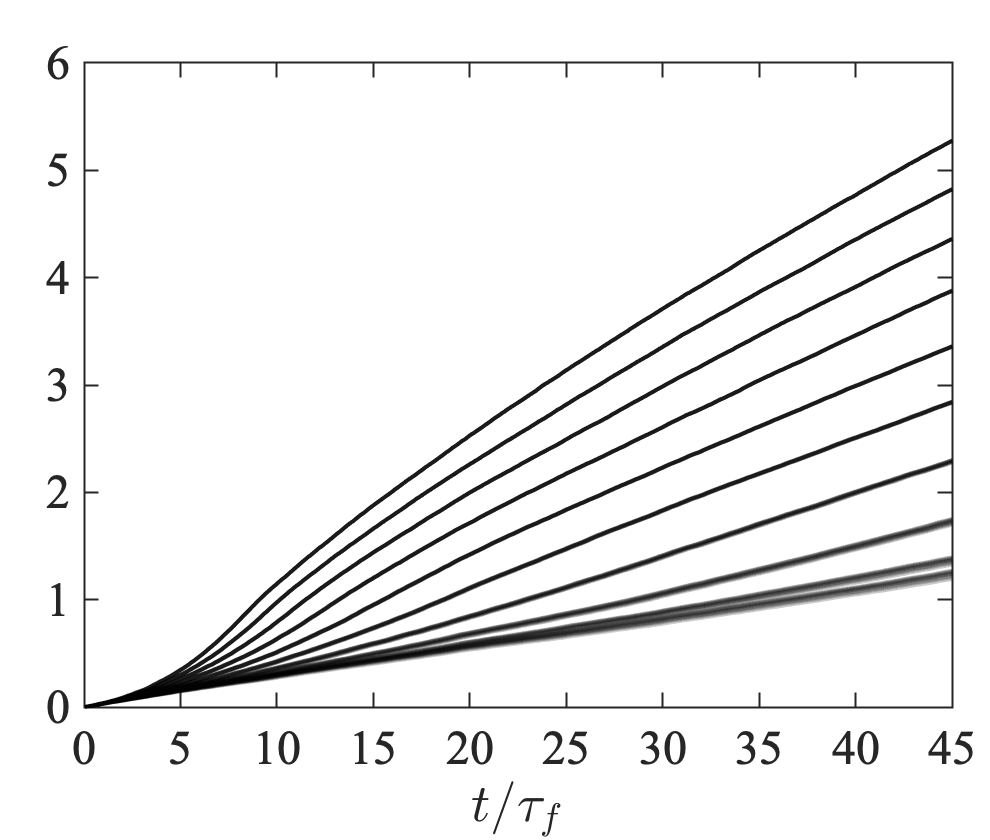}\label{dep-c}}\quad
      \subfloat[]{\includegraphics[width=.34\textwidth,valign=b]{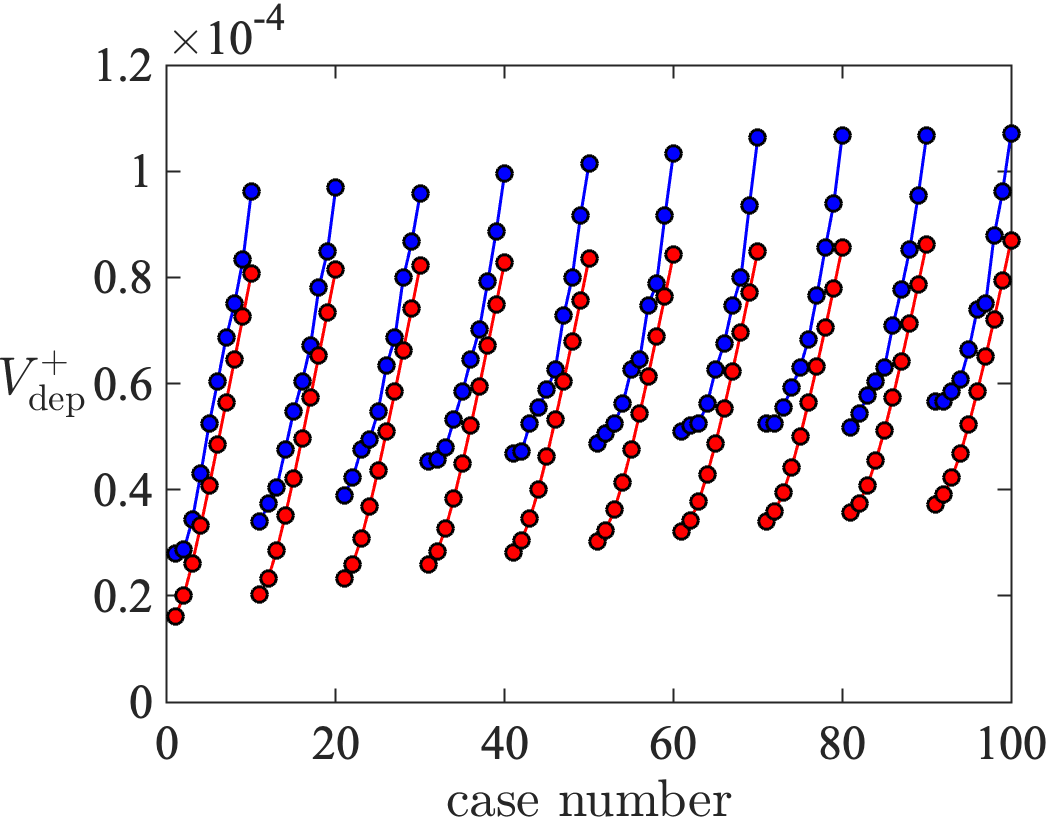}\label{dep-d}}
   \subfloat[]{\includegraphics[width=.32\textwidth,valign=b]{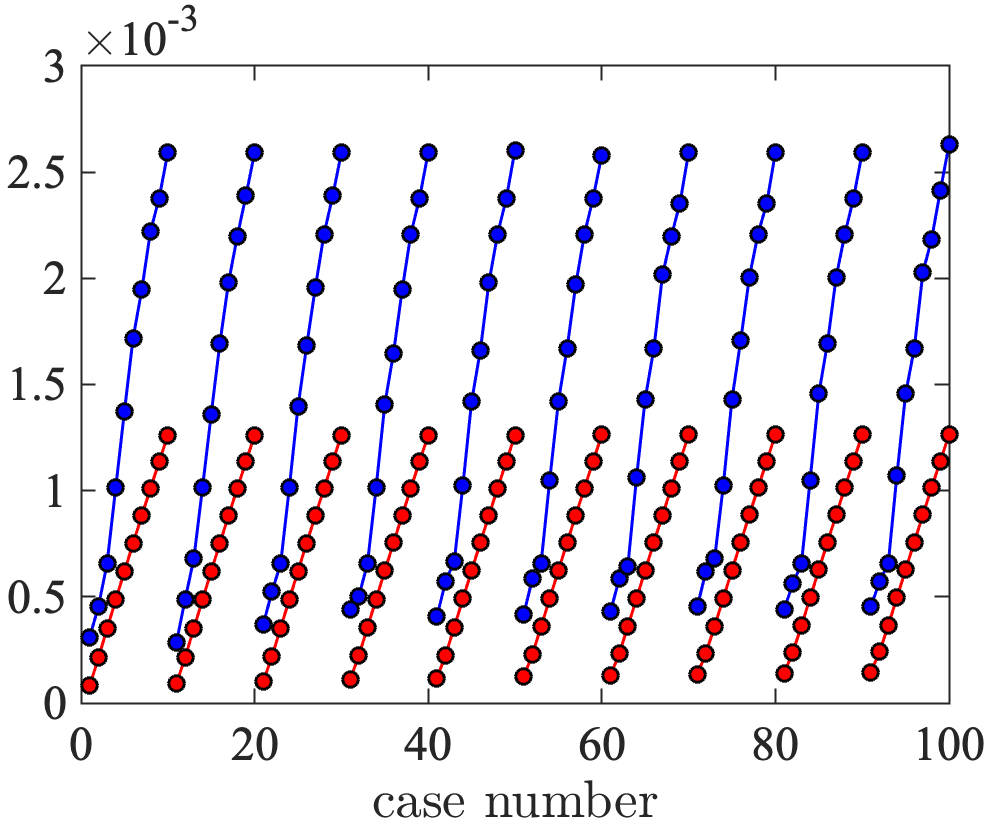}\label{dep-e}}
    \subfloat[]{\includegraphics[width=.33\textwidth,valign=b]{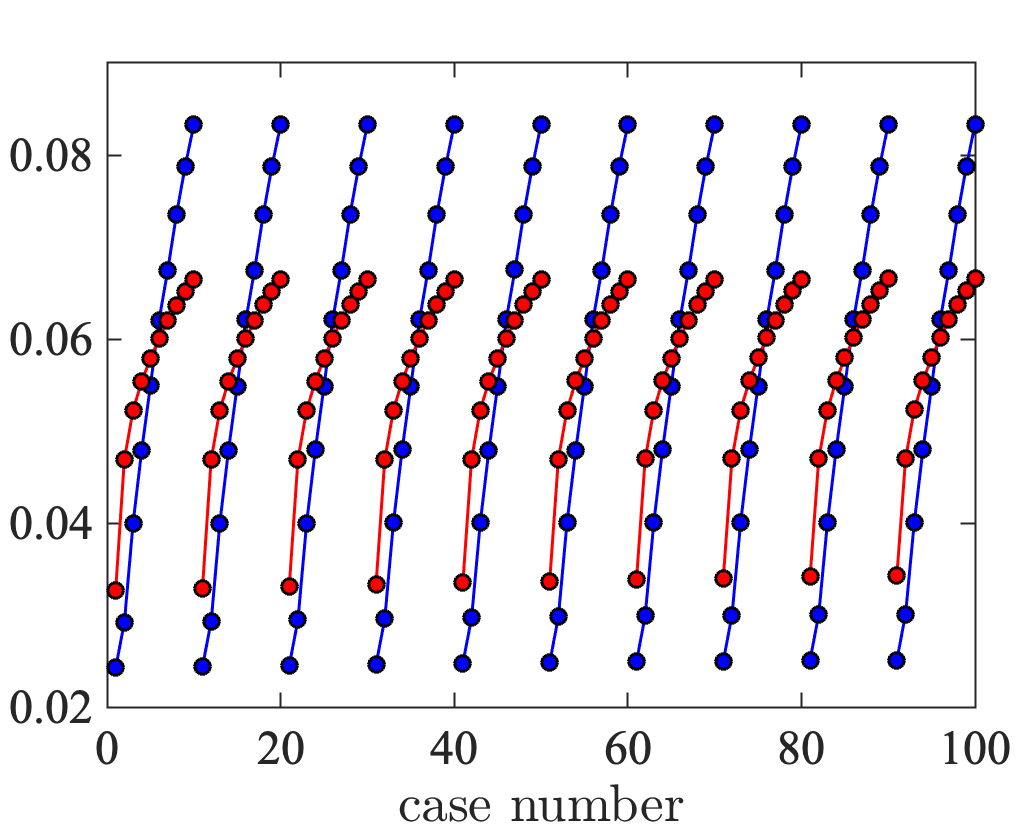}\label{dep-f}}
  \caption{(a)-(c) Percent of the total number of particles deposited from 100 DNS simulations. The slopes are used to determine the deposition rates. (d)-(f) Non-dimensional deposition velocities predicted by DNS ($\color{blue}{-\color{blue}{\bullet}-}$) and the one-dimensional Eulerian model ($\color{red}{-\color{red}{\bullet}-}$) over the 100 runs. Each line spans the range of $\xi$ with increasing $\zeta$ from left to right.}  
  \label{fig:dep_compare}
\end{figure}

Three particle sizes ($\tau_p^{+} = 0.1, 1, {\rm and\ }10$) are considered, corresponding to each deposition regime.
The deposition rates obtained using DNS are measured from the rates at which the number of deposited particles $N_{\rm dep}$ increases in time. As shown in Figs.~\ref{dep-a}--\ref{dep-c}, $N_{\rm dep}$ grows linearly in time after a short initial transient ($t/\tau_f<10$), beyond which the deposition rate can be uniquely determined. The van der Waals interaction is seen to have comparable effects on the deposition rate to the electrostatic force for small particles ($\tau_p^{+} = 0.1$), resulting in a wide spread of the deposition curves. However, as the electrostatic contribution quickly surpasses the van der Waals for larger particles, the variation in van der Waals strength ($\zeta$) exhibits little effect on the deposition except when $\xi\approx 0$.

The computed deposition rate in terms of non-dimensional deposition velocity $V_{\rm dep}^{+}$ is compared with the one-dimensional Eulerian model predictions in Figs.~\ref{dep-d}--\ref{dep-f}. The one-dimensional model is seen to under-predict $V_{\rm dep}^{+}$ for all three particle sizes, especially for mid-size particles for which clustering due to turbulence is expected to be most significant. Recall that the one-dimensional model only accounts for particle-wall interactions and the particle distribution is assumed to be only a function of wall distance. Thus, spatial inhomogeneities due to preferential concentration and low-speed streak regions along the wall are neglected. These inhomogeneities amplify particle-particle interactions and collision rates, which may potentially contribute to particle deposition. Nevertheless, the general trends of $V_{\rm dep}^{+}$ predicted by the one-dimensional model and three-dimensional DNS are in reasonable agreement. The high correlation between the two models 
suggests that
MFMC UQ, presented in the following section, will likely provide substantial benefits.

\section{Uncertainty quantification framework}\label{sec:mfmc}

Our UQ task is to compute the uncertainty on model output QoI $V_{\rm dep}^{+}$ resulting from the uncertainty of  model inputs $\xi$ and $\zeta$. 
We employ MFMC to perform this computation~\cite{peherstorfer2016optimal,peherstorfer2018survey}.
MFMC is a MC method that makes use of multiple models, exploiting their tradeoffs of fidelity versus computational cost by optimizing the work distribution across the different models
such that the MC error is minimized for a given computational budget.
MFMC leverages low-fidelity models to accelerate uncertainty propagation while guaranteeing unbiased estimators by occasionally recoursing to a high-fidelity model. In this section, both MC and MFMC estimation are summarized. Details on the various models employed are provided in~\ref{sec:hf_model} and~\ref{sec:lf_model}. The application of MFMC for UQ of particle deposition rate in turbulent pipe flow is given in subsequent sections.

\subsection{
{Monte Carlo estimation} 
}
The high-fidelity model (DNS) is denoted as
\begin{equation}
f^{(1)} : \mathcal{X} \rightarrow \mathcal{Y},
\end{equation}
where $\boldsymbol{x} \in \mathcal{X}$ is the input and $y=f^{(1)}(\boldsymbol{x}) \in \mathcal{Y}$ is the output. 
Here, the input parameters $\xi \in [0, \,0.1]$ and $\zeta \in [0,\,1]$ together form a random vector $\boldsymbol{X}$ associated with a probability distribution, which in this case is the uniform distribution discussed earlier. The output $Y=V_{\rm dep}^{+} = f^{(1)}(\boldsymbol{X})$ is therefore a random variable whose distribution is induced by the distribution of $\boldsymbol{X}$. Our goal is to
characterize the distribution of $f^{(1)}(\boldsymbol{X})$, or more precisely, find the pushforward probability measure for $\boldsymbol{X}$ through the mapping $f^{(1)}$.

Typically, we are interested in expectations of various functions with respect to this distribution (e.g., $\mathbb{E}[h(Y)]$), which can be estimated using MC.
For clarity, we present MC below 
in terms of the expected value 
$\mathbb{E}[Y]$
(i.e. the mean), but 
the results are valid for general expectations with $h(Y)$ as well.
%
The MC estimator for
\begin{equation}
S=
\mathbb{E}[Y] = \mathbb{E}[f^{(1)}(\boldsymbol{X})].
\end{equation}
is 
\begin{equation}
\overline{y}_{m}=\frac{1}{m} \sum_{j=1}^{m} f^{(1)}\left(\boldsymbol{x}_{j}\right)
\end{equation}
where $\boldsymbol{x}_{j}$ are independent and identically distributed (i.i.d.) samples of $\boldsymbol{X}$.
The MC estimator is unbiased ($\mathbb{E}\left[\overline{y}_{m}\right]=S$), and so its 
mean-squared error (MSE) also equals to the estimator variance
\begin{equation}
\mathrm{MSE}\left(\overline{y}_{m}\right)=
\mathbb{E}\left[\left(\overline{y}_{m}-S\right)^{2}\right] =\operatorname{Var}\left[\overline{y}_{m}\right]= \frac{\operatorname{Var}[f^{(1)}(\boldsymbol{X})]}{m}.
\label{eq:MSE}
\end{equation}
{Under the reasonable assumption that the computational cost of each high-fidelity model run is approximately the same, }
the total computational cost of the MC estimator $\overline{y}_{m}$ is then
\begin{equation}
c\left(\overline{y}_{m}\right)=w m,
\end{equation}
where $w$ is the unit cost of each high-fidelity evaluation. Depending on the variance  of the output variable ($\operatorname{Var}[f^{(1)}(\boldsymbol{X})]$), a large number of high-fidelity model evaluations ($m$) may be necessary to obtain an MC estimate with an acceptable MSE. If the high-fidelity model is expensive to evaluate, then MC estimation can become intractable.

\subsection{{Multi-fidelity Monte Carlo} 
}\label{sec:mfmc_mean}

MFMC 
makes use of low-fidelity models, in addition to the high-fidelity model,
with the aim to minimize the MC estimator MSE by optimizing the number of evaluations for each model under a given total computational budget. Suppose we have $k$ models sorted in
descending
fidelity $f^{(1)},\ldots,f^{(k)}$ ($f^{(1)}$ is still our targeted high-fidelity model), each evaluated $m_k$ times ($0<m_{1} \leq m_{2} \leq \cdots \leq m_{k}$), then their individual MC estimators are defined by
\begin{equation}
\overline{y}_{m_{i}}^{(i)}=\frac{1}{m_{i}} \sum_{j=1}^{m_{i}} f^{(i)}\left(\boldsymbol{x}_{j}\right) \qquad i=1,\ldots,k. \end{equation}
The MFMC estimator for the highest fidelity model is formed using the control variates technique, through a linear combination of these MC estimators:
\begin{equation}\label{eq:MFMC_mean}
\hat{S}=\overline{y}_{m_{1}}+\sum_{i=2}^{k} \alpha_{i}\left(\overline{y}_{m_{i}}^{(i)}-\overline{y}_{m_{i-1}}^{(i)}\right).
\end{equation}
MFMC seeks the set of coefficients $\alpha_i$ and number of evaluations $m_i$ 
that minimizes $\operatorname{Var}[\hat{S}]$ subject to the total cost constraint
\begin{equation}
c(\hat{S})=\sum_{i=1}^{k} m_{i} w_{i} \le p
\end{equation} 
where $w_{i}$ is the unit cost of model $f^{(i)}$ and $p$ is the given total computational budget. The optimization procedure requires the correlation coefficients between each low-fidelity model to the targeted high-fidelity model, given by
\begin{equation}
\rho_{1,i}=\frac{\operatorname{Cov}\left[f^{(1)}(\boldsymbol{X}), f^{(i)}(\boldsymbol{X})\right]}{\sqrt{\operatorname{Var}\left[f^{(1)}(\boldsymbol{X})\right] \operatorname{Var}\left[f^{(i)}(\boldsymbol{X})\right]}}, \quad i=2, \ldots, k.
\end{equation}
An analytical solution \citep{peherstorfer2016optimal} to this optimization problem yields, for $i=2, \ldots, k$:
\begin{equation}\label{eq:opt1}
\alpha_{i}^{*}=\frac{\rho_{1,i} \sigma_{1}}{\sigma_{i}}
\end{equation}
\begin{equation}\label{eq:opt2}
r_{i}^{*}=\sqrt{\frac{w_{1}\left(\rho_{1,i}^{2}-\rho_{1,i+1}^{2}\right)}{w_{i}\left(1-\rho_{1,2}^{2}\right)}}
\end{equation}
\begin{equation}\label{eq:opt3}
m_i^{*} = m_1^{*}r_i^{*}
\end{equation}
where $m_1^*=p/(\boldsymbol{w}^T \boldsymbol{r}^*)$ and $\sigma_{i}=\sqrt{\operatorname{Var}\left[f^{(i)}(\boldsymbol{X})\right]}$ is the standard deviation of model $f^{(i)}(\boldsymbol{X})$. A mild condition on the model costs is required for this analytical solution to exist: 
\begin{equation}\label{eq:mf-precond}
\frac{w_{i-1}}{w_{i}}>\frac{\rho_{1, i-1}^{2}-\rho_{1, i}^{2}}{\rho_{1, i}^{2}-\rho_{1, i+1}^{2}}
\end{equation}
for $i=2, \ldots, k$ with $\rho_{1,k+1}=1$. We will show in \S~\ref{sec:mfmc_verification} that Eq.~\eqref{eq:mf-precond} can be easily satisfied for our purposes.

\subsection{Multi-fidelity Monte Carlo for global sensitivity analysis}\label{sec:mfmc_sobol}


Variance-based GSA provides a quantitative measure of how uncertainty from each model input contributes to the overall uncertainty in the model output QoI, through the Sobol' indices. The two most commonly used Sobol' indices are the main-effect index and total-effect index. The main-effect index measures variance contribution solely
due to the $j$-th input:
\begin{equation}
s_j \equiv \frac{V_j}{{\rm Var}(Y)} = \frac{\operatorname{Var}_{\boldsymbol{X}_{j}}\left(\mathbb{E}_{{\boldsymbol{X}}_{\sim j}}\left[Y \mid \boldsymbol{X}_{j}\right]\right)}{{\rm Var}(Y)},
\end{equation}
where $Y$ is the output QoI
and $\boldsymbol{X}_{\sim j}$ denotes the set of all input variables except the $\boldsymbol{X}_{j}$ component. 
The total-effect index measures variance contributions from \emph{all}
terms that involve the $j$-th parameter, including higher-order interaction effects among multiple inputs that involve the $j$-th parameter:
\begin{equation}
s_j^{t} \equiv \frac{T_j}{{\rm Var}(Y)} = \frac{\mathbb{E}_{{\boldsymbol{X}}_{\sim j}}\left[\operatorname{Var}_{\boldsymbol{X}_{j}}\left(Y \mid {\boldsymbol{X}}_{\sim j}\right)\right]}{\operatorname{Var}(Y)}.
\end{equation}
Thus, the difference between $s_j$ and $s_j^t$ offers a sense about the presence of interaction effects without explicitly computing them.

While these Sobol' indices can be estimated with single-fidelity MC sampling \citep{Sobol1990,Jansen1999,Saltelli1999,Saltelli2002},
\citet{qian2018multifidelity} has recently proposed
a MFMC estimator similar to the form of Eq.~\eqref{eq:MFMC_mean}.
The MFMC estimate of the Sobol' indices are $s_j = \hat{V}_{j, \mathrm{mf}}/{{\rm Var}(Y)}$ and $s_j^{t} = \hat{T}_{j, \mathrm{mf}}/{{\rm Var}(Y)}$, where
\begin{equation}
\hat{V}_{j, \mathrm{mf}}=\hat{V}_{j, m_{1}}^{(1)}+\sum_{i=2}^{k} \alpha_{i}\left(\hat{V}_{j, m_{i}}^{(i)}-\hat{V}_{j, m_{i-1}}^{(i)}\right)
\end{equation}
and 
\begin{equation}
\hat{T}_{j, \mathrm{mf}}=\hat{T}_{j, m_{1}}^{(1)}+\sum_{i=2}^{k} \alpha_{i}\left(\hat{T}_{j, m_{i}}^{(i)}-\hat{T}_{j, m_{i-1}}^{(i)}\right).
\end{equation}
We further adopt an unbiased MC estimator for $\hat{V}_{j,m_i}^{(i)}$ and $\hat{T}_{j,m_i}^{(i)}$ proposed by~\citet{owen2013variance}:
\begin{equation}
\hat{V}_{j,m_i}^{(i)}=\frac{2 m_i}{2 m_i-1}\left(\frac{1}{m_i} \sum_{i=1}^{m_i} f^{(i)}\left(\bm{x}_{i}\right) f^{(i)}\left(\bm{z}_{i}^{(j)}\right)-\left(\frac{\hat{E}+\hat{E}^{\prime}}{2}\right)^{2}+\frac{\hat{V}+\hat{V}^{\prime}}{4 m_i}\right)
\end{equation}
and 
\begin{equation}
\hat{T}_{j,m_i}^{(i)}=\frac{1}{2 m_i} \sum_{i=1}^{m_i}\left(f^{(i)}\left(\bm{x}_{i}^{\prime}\right)-f^{(i)}\left(\bm{z}_{i}^{(j)}\right)\right)^{2},
\end{equation}
where $\bm{z}_{i}^{(j)} = (\bm{x}_i'(1),\dots,\bm{x}_i'(j-1)),\bm{x}_i(j),\bm{x}_i'(j+1),\dots,\bm{x}_i'(d))$ with $d$ being the number of input parameters, $\hat{E}$, $\hat{E}'$ and $\hat{V}$, $\hat{V}'$ are the sample means and variances estimated using two sets of $m_i$ independent realizations of input $\boldsymbol{X}$ respectively ($\boldsymbol{x}_{1}, \ldots, \boldsymbol{x}_{m_i}$ and $\boldsymbol{x}'_{1}, \ldots, \boldsymbol{x}'_{m_i}$). 
%
Furthermore, \citet{qian2018multifidelity} demonstrated that the same optimization procedures (Eqs.~\eqref{eq:opt1}--\eqref{eq:opt3}), namely same $\alpha_i$ and $m_i$ for $i = 2, \dots, k$, can be used for the MFMC estimates of the
Sobol' indices 
without sacrificing the performance of the algorithm.

\section{Uncertainty quantification and sensitivity analysis of particle deposition}


In this section, we conduct GSA for the deposition rate by estimating their main-effect Sobol' indices using the MFMC method. 
MFMC aims to minimize the variance of the MC estimator for the Sobol' indices under a given computational budget, 
or equivalently, use minimal computational resources to achieve an MC error requirement. 
The computational costs, model statistics and correlations found from the 100 tensorized runs in Section~\ref{ss:comparison} are summarized in Table~\ref{table:mf-stats}.
\begin{table}
 \begin{center}
\begin{tabular}{l|lllll}
\hline $\tau_p^{+}$ & Model & $\mu_{i}$ & $\sigma_{i}$ & $\rho_{1 i}$ & $w_{i}$ \\
\hline \multirow{2}{*}{0.1} & $f^{(1)}={\rm DNS}$ & $6.73\times 10^{-5}$ & $1.91\times 10^{-5}$ & 1 & 1 \\
& $f^{(2)}={\rm 1D\ Eulerian}$ & $5.28\times 10^{-5}$ & $1.88\times 10^{-5}$ & $0.989$ & $1/2880$ \\
\hline \multirow{2}{*}{1} & $f^{(1)}={\rm DNS}$ & $1.52\times 10^{-3}$ & $7.63\times 10^{-4}$ & 1 & 1 \\
& $f^{(2)}={\rm 1D\ Eulerian}$ & $7.02\times 10^{-4}$ & $3.72\times 10^{-4}$ & $0.995$ & $1/2880$ \\
\hline \multirow{2}{*}{10} & $f^{(1)}={\rm DNS}$ & $5.74\times 10^{-2}$ & $1.96\times 10^{-2}$ & 1 & 1 \\
& $f^{(2)}={\rm 1D\ Eulerian}$ & $5.68\times 10^{-2}$ & $1.01\times 10^{-2}$ & $0.937$ & $1/2880$ \\
\hline
\end{tabular}
\caption{Prior estimates of model mean ($\mu_i$), standard deviation ($\sigma_i$) of the deposition rate $V_{\rm dep}^{+}$, correlation coefficients ($\rho_{1i}$), and costs ($w_i$) used for multi-fidelity sensitivity analysis for three different particle sizes.}
   \label{table:mf-stats}
 \end{center}
\end{table} 
For demonstration purposes, 
these model statistics are computed using all 100 runs before feeding them into the multi-fidelity optimization (Eqs.~\eqref{eq:opt1}--\eqref{eq:opt3}). In practice, however, a pilot set with a small number of samples is typically used to estimate these statistics. \citet{peherstorfer2016optimal} demonstrated that the MFMC
run allocations
are fairly insensitive to the these prior estimates of model statistics; we indeed observed similar results with fewer runs. 
Furthermore, data from such pilot runs can be reused 
for the main MFMC computations
and therefore would not be wasted.

The computational budget is measured in terms of CPU hours and is normalized such that the cost of one converged DNS run is unity, which is estimated to be 256 CPU hours. The average cost of each one-dimensional model is only $1/2880$. As shown in Table~\ref{table:mf-stats}, the mean ($\mu_i$) and standard deviation ($\sigma_i$) of $V_{\rm dep}^{+}$ are comparable between these two models, resulting in high correlation coefficients ($\rho_{12}$) for all three cases. It is interesting to note that the $\rho_{12}$ for $\tau_p^{+}=1$ is the highest despite the relatively large discrepancy between the DNS and one-dimensional model statistics. This is because by definition, $\rho_{12}$ only reflects the similarity of model prediction trends and is agnostic to the absolute values of its mean or standard deviation. The difference in these model statistics is later compensated by the control variate coefficients $\alpha_i$.
Unlike mean or variance estimates, Sobol' index estimates require $(d+2)$ model evaluations for each Monte Carlo sample, and so an effective budget $p_{\rm eff}=p/(d+2)$ introduced by \citet{qian2018multifidelity} is used for the optimization instead of $p$. 

To fully leverage the 100 high-fidelity DNS runs already completed for the tensorized study in \S~\ref{ss:comparison} and to illustrate the convergence of our MFMC approach, we will restrict our MFMC samples to be selected only from this pre-produced set. One may view such setup as a discrete measure approximation to the continuous uniform distributions on $\xi$ and $\zeta$, and the MFMC estimator will thus converge to this discrete approximate expectation in the limit of infinite samples. With samples only generated from these 100 locations, it is also possible for sample points to repeat, especially with larger $m_i$. A repeated sample would not incur any additional computational cost, since it has been previously computed. When this happens, the assumption of constant unit cost per simulation breaks down, and the MFMC solution in \S~\ref{sec:mfmc_mean} would be, strictly speaking, sub-optimal. However, such ``saturation'' is unlikely to occur with low $m_i$, especially for the high-fidelity model where the computational cost usually dominates; this is indeed our case here as we will show. 
In practice where such tensorized, discrete measure approximation is typically unavailable, one would still deploy random sampling in the continuous input space. Nevertheless, the unbiased MFMC formulation using the control variates technique and optimization procedure described in \S~\ref{sec:mfmc} still hold regardless of sampling methods.

\subsection{Validation of the MFMC framework for particle deposition}\label{sec:mfmc_verification}

Prior to preforming the actual uncertainty quantification, it is important to ensure the MFMC precondition (Eq.~\eqref{eq:mf-precond}) is met and the MFMC estimator is superior to the classic MC estimator for the two models considered in this study. To assess the performance of an estimator, the 
MSE introduced in
Eq.~\eqref{eq:MSE}
is typically used and estimated using samples:
\begin{equation}\label{eq:mse}
{\rm MSE}\left(\hat{S}\right) = \frac{1}{m} \sum_{i=1}^{m}\left(S_{i}-\hat{S}\right)^{2},
\end{equation}
where $S_i = f^{(i)}(\bm{x})$ and $\hat{S}$ is the estimator prediction. In the case of two models, \citet{peherstorfer2016optimal} derived an analytical estimate of the variance reduction achieved by the MFMC estimator compared to the classic MC estimator given as 
\begin{equation}
\gamma \equiv \frac{{\rm MSE}\left(\hat{S}_{\rm mf}\right)}{{\rm MSE}\Big( \overline{y}_{m} \Big)}=\left(\sqrt{1-\rho_{12}^2} + \rho_{12} \sqrt{w_2/w_1}\right)^2 
\end{equation}
where $\gamma$ is the variance reduction ratio, $\hat{S}_{\rm mf}$ denotes the MFMC estimator.

\begin{figure}[h!]
  \centering
  {\includegraphics[width=.6\textwidth]{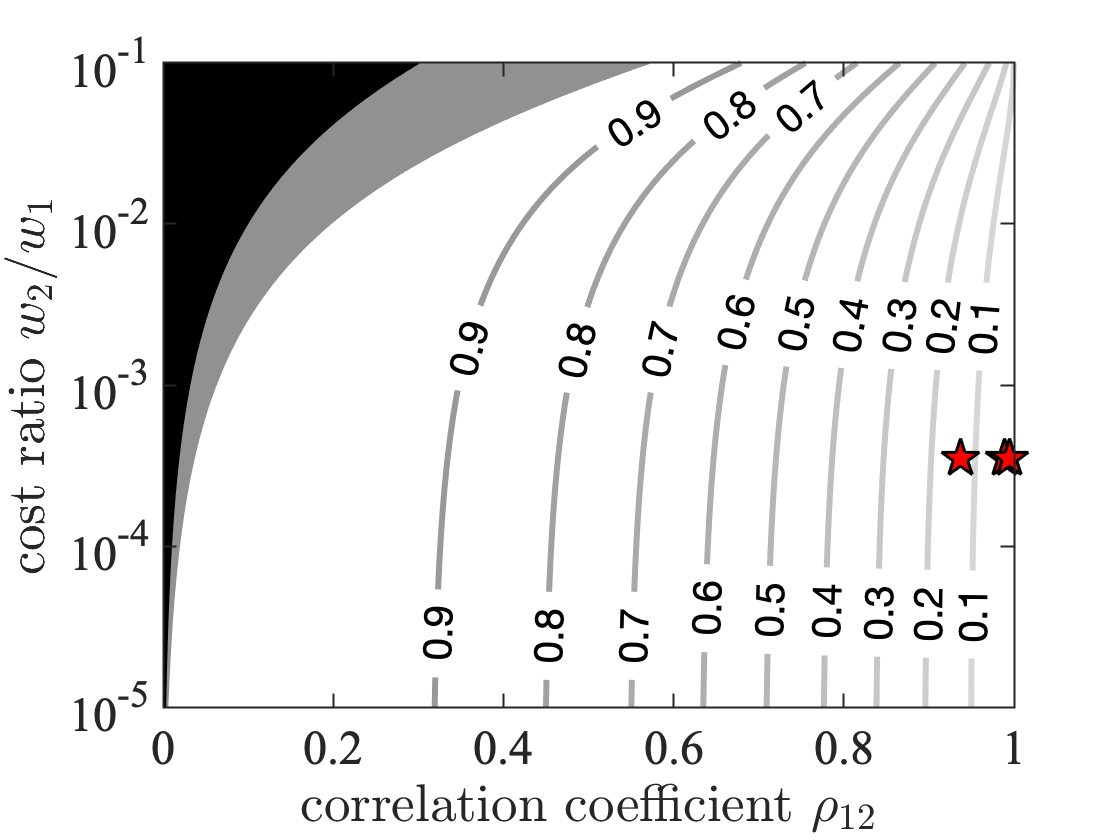}}
  \caption[Contours of variance reduction ratio as a function of model costs and correlation coefficients.]{Contours of variance reduction ratio $\gamma$ as a function of model cost ratio $w_2/w_1$ and correlation coefficient $\rho_{12}$. The black region shows where the precondition (Eq.~\eqref{eq:mf-precond}) is violated. The gray region represents where the MFMC estimators fail to reduce the variance of the QoIs compared to classic MC (i.e., $\gamma>1$). $\gamma=0.028, 0.014,$ and 0.135 for $\tau_p^{+}=0.1,1,$ and 10 respectively, denoted by red stars. }  
  \label{fig:regime-mf}
\end{figure}

Figure~\ref{fig:regime-mf} shows the contour plot of $\gamma$ for $\rho_{12} \in [0,1]$ and $w_2/w_1 \in [10^{-5},10^{-1}]$. The multi-fidelity approach fails when $\gamma>1$ or the precondition (Eq.~\eqref{eq:mf-precond}) is not met. However, this only occurs for extreme cases, either highly uncorrelated models (e.g., $\rho_{12} <0.3$) or expensive low-order models (e.g., $w_2/w_1>10^{-2}$). The values in Table~\ref{table:mf-stats} yield $\gamma=0.028, 0.014,$ and 0.135 for $\tau_p^{+}=0.1,1,$ and 10 respectively, confirming that the MFMC estimator will significantly improve MSE using these two models.

\subsection{Uncertainty quantification of the deposition rate}
Using the prior estimates of the model statistics provided in Table~\ref{table:mf-stats}, the effective budget $p_{\rm eff}$ is then distributed among our two models of consideration. Table~\ref{table:mf-evals} summarizes the number of evaluations, $m_k$, and the control variate coefficients $\alpha_k$ assigned by the MFMC approach for $p=36$ and $72$. The same set of $m_k$ and $\alpha_k$ is then used to estimate the mean, variance, and sensitivity indices. As expected, more evaluations are performed for the one-dimensional model when it is more correlated with the DNS predictions (e.g., $\tau_p^{+}=1$), and a higher coefficient $\alpha_k$ is required when the one-dimensional model under-predicts $\sigma_k$ (e.g., $\tau_p^{+}=1$ and $10$). The classic MC method using only high-fidelity (DNS) and only low-fidelity (one-dimensional Eulerian) models are also included for subsequent comparisons. 

\begin{table}
 \begin{center}
\begin{tabular}{c|c|cc|cc|cc|cc}
\hline \multirow{2}{*}{$\tau_p^{+}$} & \multirow{2}{*}{Model} &\multicolumn{2}{|c|} {  MC ($p=36$) }& \multicolumn{2}{|c|} { MF ($p=36$) } & \multicolumn{2}{|c|} {  MC ($p=72$) } & \multicolumn{2}{c} { MF ($p=72$) } \\
& & \multicolumn{2}{|c|} {  $m_{k}$ } & $m_{k}$ & $\alpha_{k}$ & \multicolumn{2}{|c|} {  $m_{k}$ } & $m_{k}$ & $\alpha_{k}$ \\
\hline \multirow{2}{*}{0.1} & $f^{(1)}$ & 9 & $-$& 8 & 1 & 18 & $-$ &16 & 1 \\
& $f^{(2)}$ & $-$ & 25920 & 2861 & $1.005$ & $-$ & 51840 & 5488 & $1.005$ \\
\hline \multirow{2}{*}{1} & $f^{(1)}$ & 9 & $-$ & 7 & 1 & 18 & $-$ & 15 & 1 \\
& $f^{(2)}$ & $-$ & 25920 & 3932 & $2.040$ & $-$ & 51840 & 7726 & $2.040$ \\
\hline \multirow{2}{*}{10} & $f^{(1)}$ & 9 & $-$ & 8 & 1 & 18 & $-$ & 17 & 1 \\
& $f^{(2)}$ & $-$ & 25920 & 1237 & $1.816$ & $-$ & 51840 & 2487 & $1.816$ \\
\hline
\end{tabular}
\caption[Number of evaluations and control variate coefficients per model.]{Number of evaluations $m_k$ and control variate coefficients $\alpha_k$ per model for MC and MF approaches with a given computational budget of $p=36$ and $p=72$.}
   \label{table:mf-evals}
 \end{center}
\end{table} 

\begin{figure}[h!]
  \centering
  \subfloat[]{\includegraphics[width=.33\textwidth]{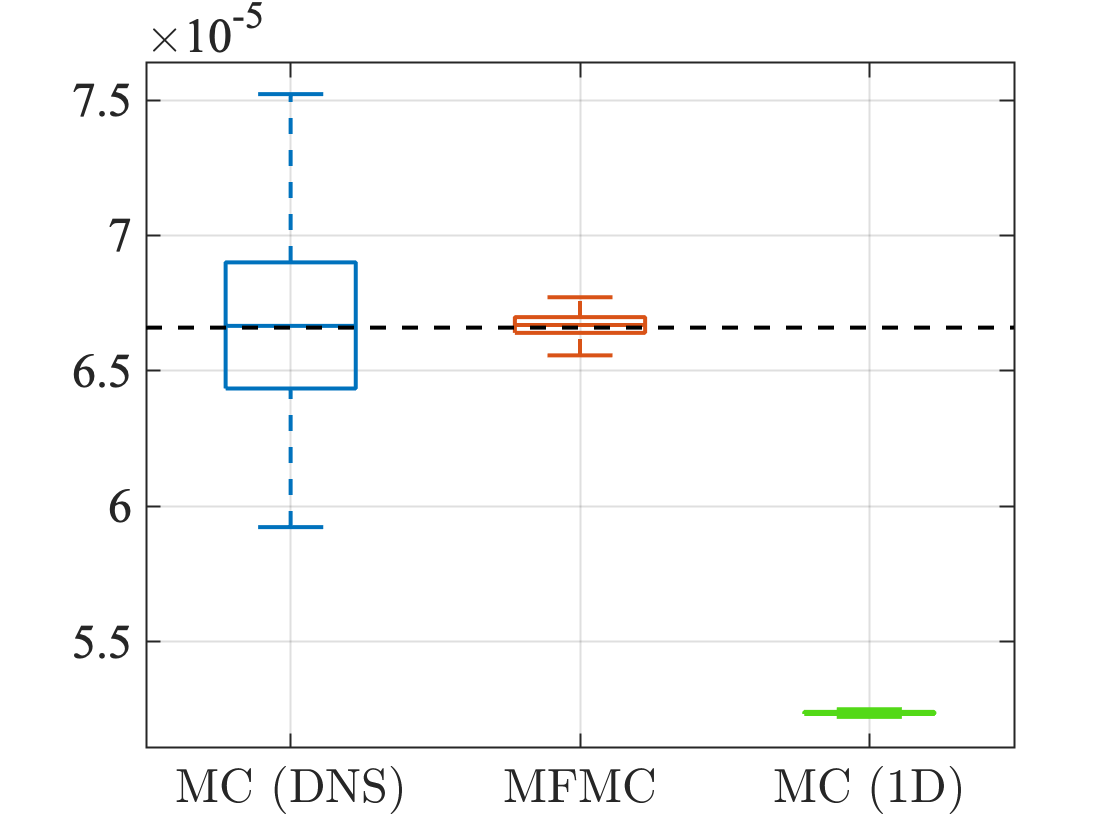}\label{mean-a}}
   \subfloat[]{\includegraphics[width=.33\textwidth]{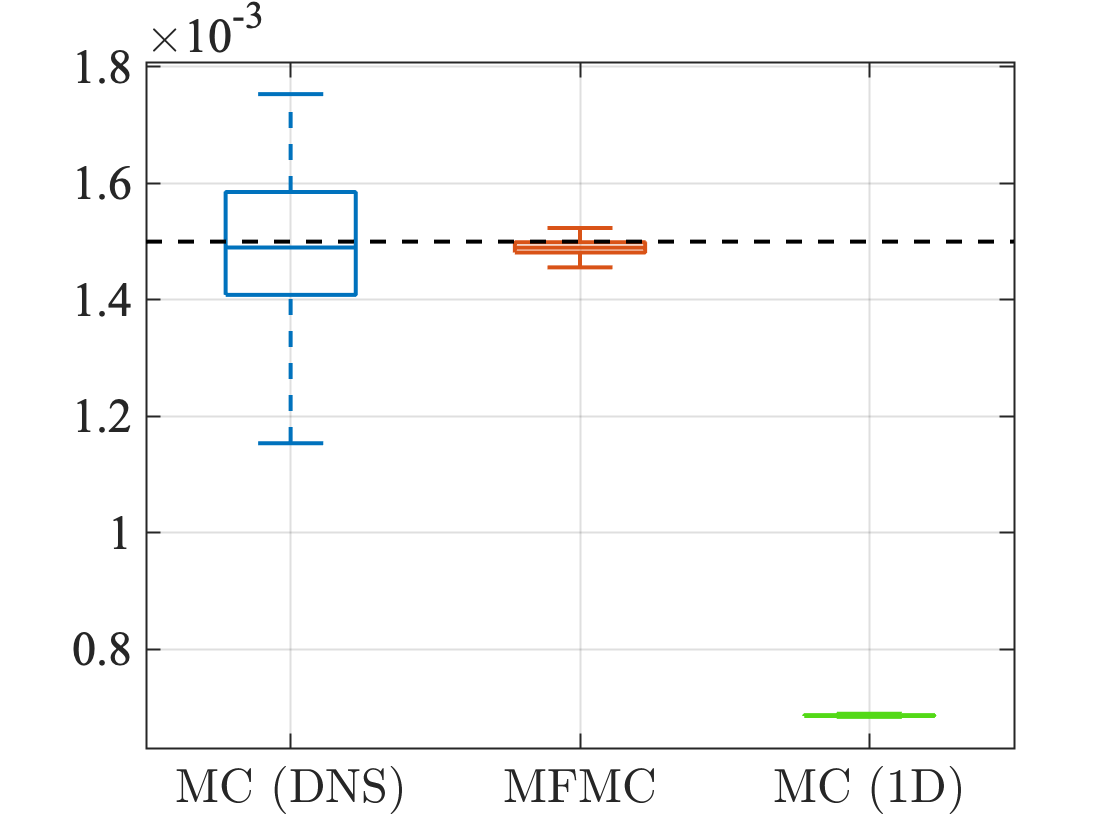}\label{mean-b}}
    \subfloat[]{\includegraphics[width=.33\textwidth]{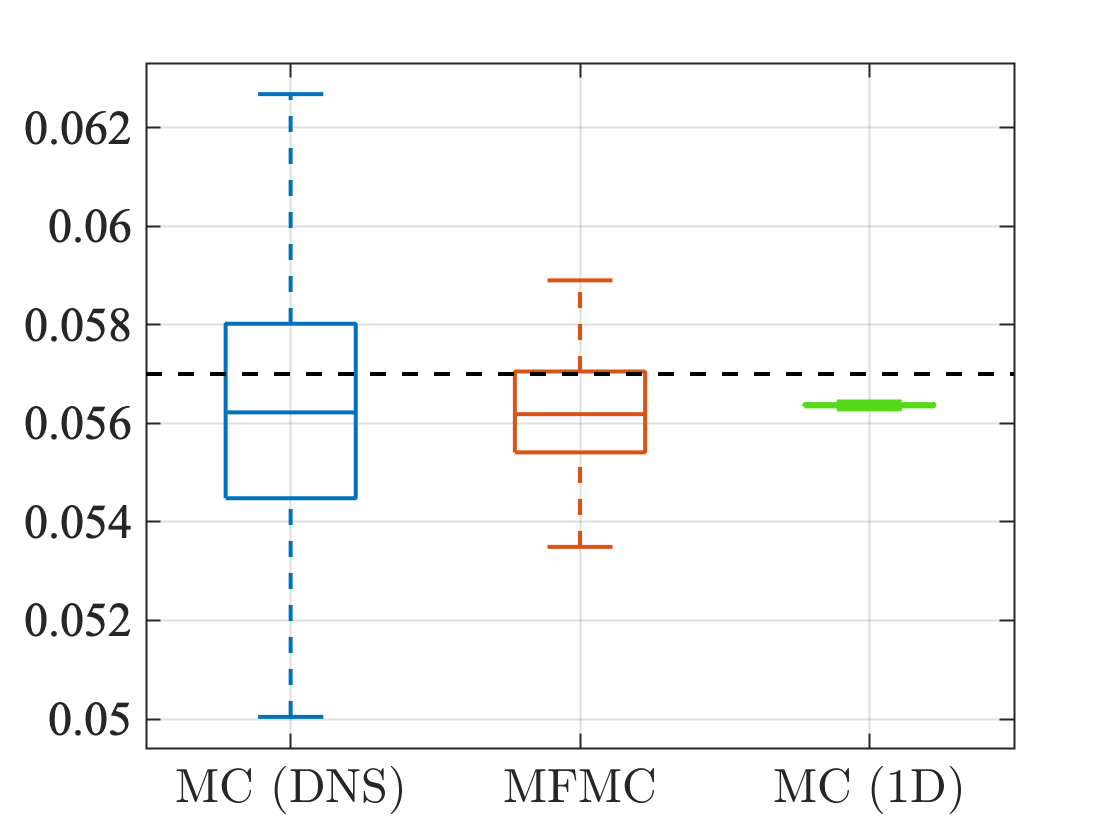}\label{mean-c}}\quad
      \subfloat[]{\includegraphics[width=.33\textwidth]{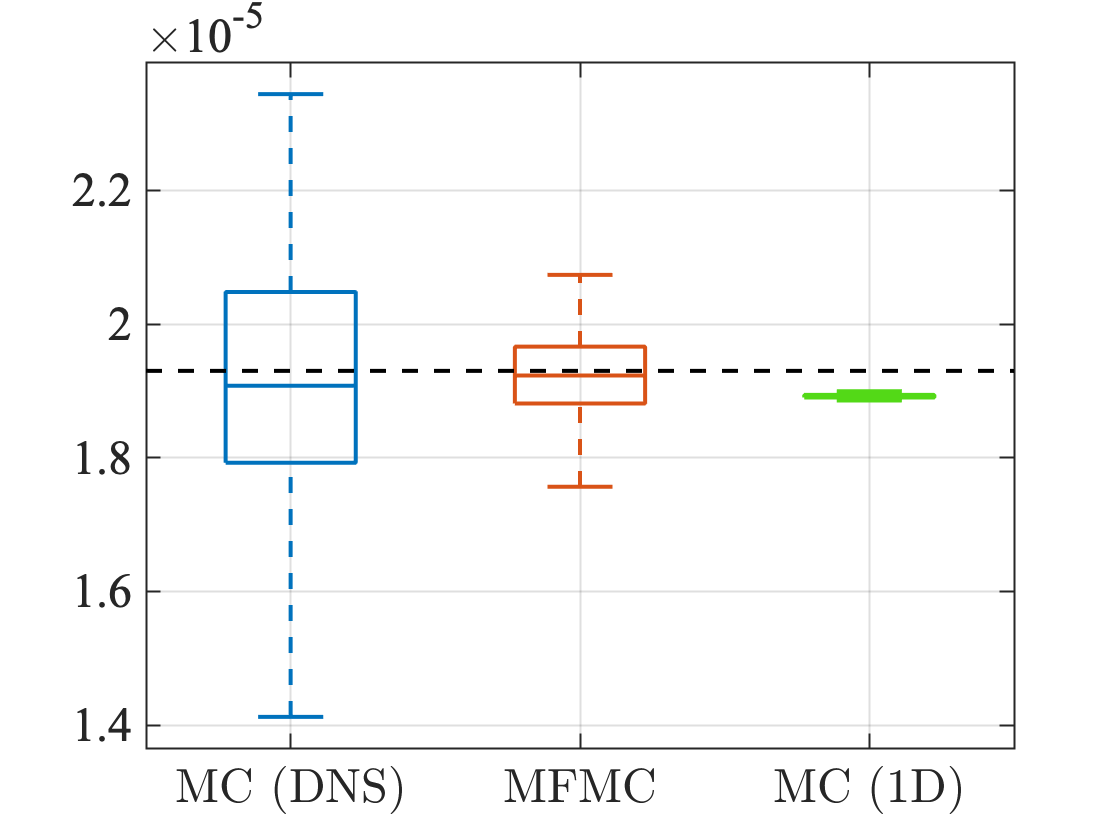}\label{var-a}}
   \subfloat[]{\includegraphics[width=.33\textwidth]{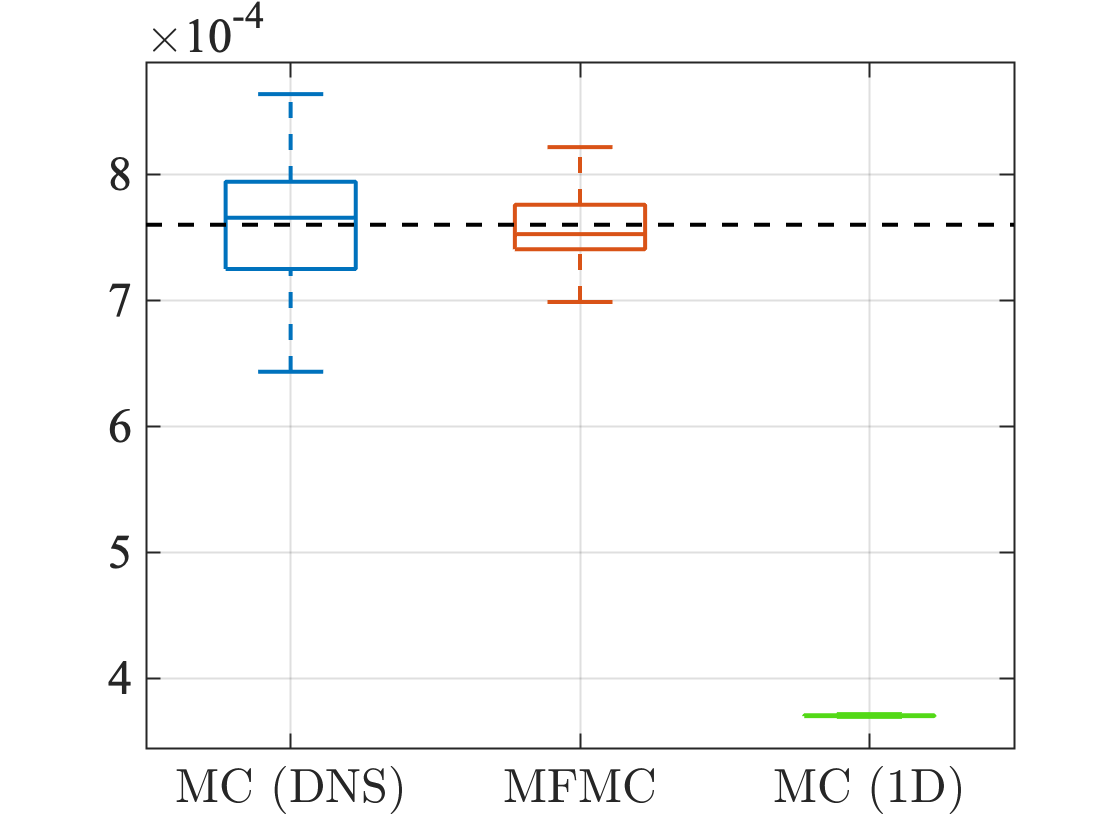}\label{var-b}}
    \subfloat[]{\includegraphics[width=.33\textwidth]{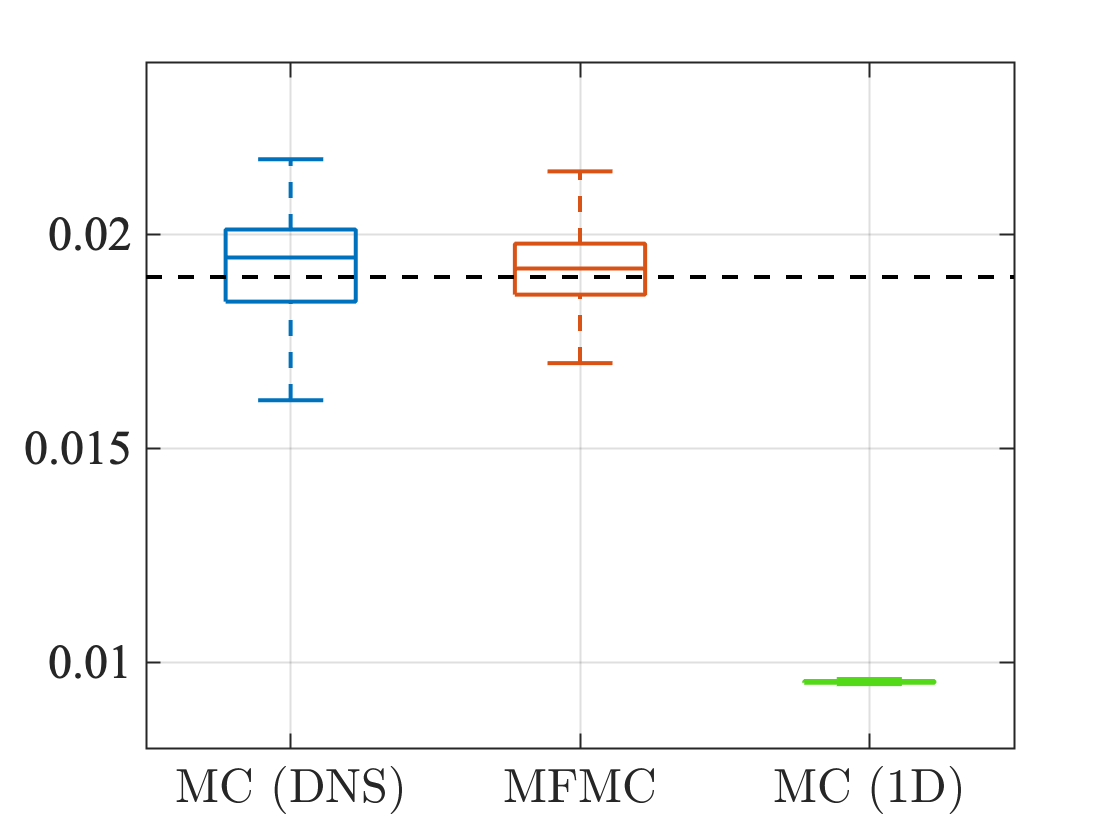}\label{var-c}}
  \caption[Box plots of 100 mean and standard deviation estimate replicates for non-dimensional deposition velocity.]{Box plots of 100 mean (top) and standard deviation (bottom) estimate replicates for $V_{\rm dep}^{+}$ with $\tau_p^{+} = 0.1, 1, {\rm \ and\ }1$ from left to right. Predictions from: MC using DNS only, 
  MFMC,
  and MC using one-dimensional model only, are shown in blue, red, and green respectively. Results reported for same budget $p=72$. Dashed line represents ``true'' values from well-converged estimates.}  
  \label{fig:mean_var}
\end{figure}

Figure~\ref{fig:mean_var} quantifies the uncertainty in mean and standard deviation of $V_{\rm dep}^{+}$ predicted by MC estimators (using only DNS or one-dimensional models) and MFMC with the same budget $p=72$. Statistics of 100 estimate replicates are shown as box plots in which the central mark indicates the median, and the bottom and top edges of the box mark the 25-th and 75-th percentiles, respectively. The whiskers extend to the lower and upper bounds of the dataset. The MFMC estimators admit notably lower variance in both the predicted mean and standard deviation compared to the high-fidelity MC estimators. Although the variance predicted by low-fidelity MC estimator is the smallest due to a larger number of evaluations, it suffers inevitable bias 
due to its modeling error (with respect to the high-fidelity model)
and therefore deviates from the ``true'' solutions, which are taken from well-converged MC estimates using $p=50000$ by randomly sampling from the 100 pre-computed DNS cases.

\begin{figure}[h!]
  \centering
  \subfloat[]{\includegraphics[width=.44\textwidth]{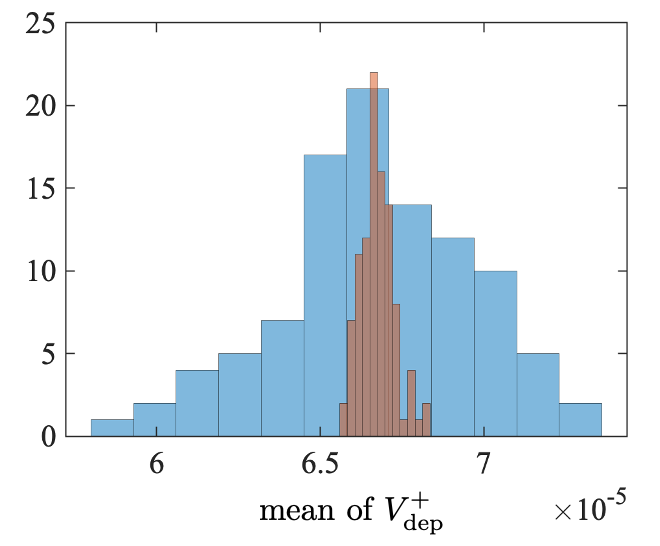}\label{bar-a}}
   \subfloat[]{\includegraphics[width=.44\textwidth]{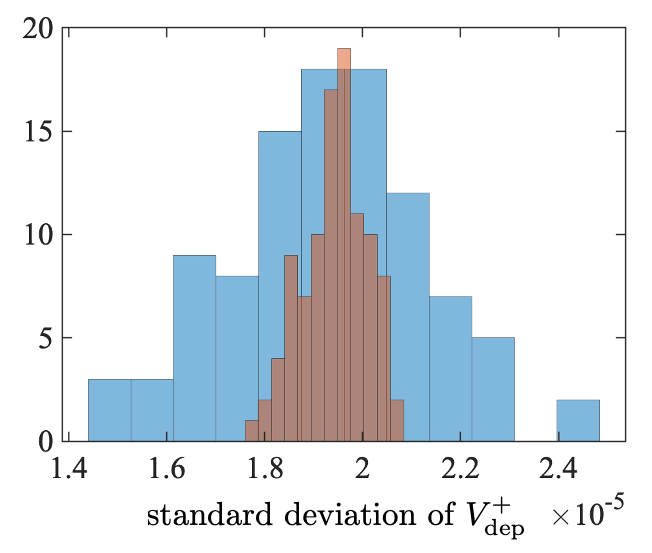}\label{bar-b}}
  \caption[Distribution of mean and standard deviation predicted by 100 replicates of MFMC and MC estimators.]{Distribution of (a) mean and (b) standard deviation predicted by 100 replicates of MFMC (red) and MC (blue) estimators with $p=72$ and $\tau_p^{+}=0.1$.}  
  \label{fig:bar}
\end{figure}

The distribution of mean and standard deviation estimates from 100 replicates of MFMC and MC estimators are shown in Fig.~\ref{fig:bar} for $p=72$ and $\tau_p^{+}=0.1$ as an example. The variations in both mean and standard deviation are significantly reduced using MFMC as indicated by the narrower distributions. The distributions are not noticeably skewed which makes the standard deviation a sufficient statistical metric. Similar reduction in estimator variations is observed for other particle sizes and computational budgets considered in Fig.~\ref{fig:mean_var}.  Note that the MFMC estimators remain unbiased even though the low-fidelity model 
produces entirely different expected values 
compared to the high-fidelity model.

\begin{figure}[h!]
  \centering
  \subfloat[]{\includegraphics[width=.33\textwidth]{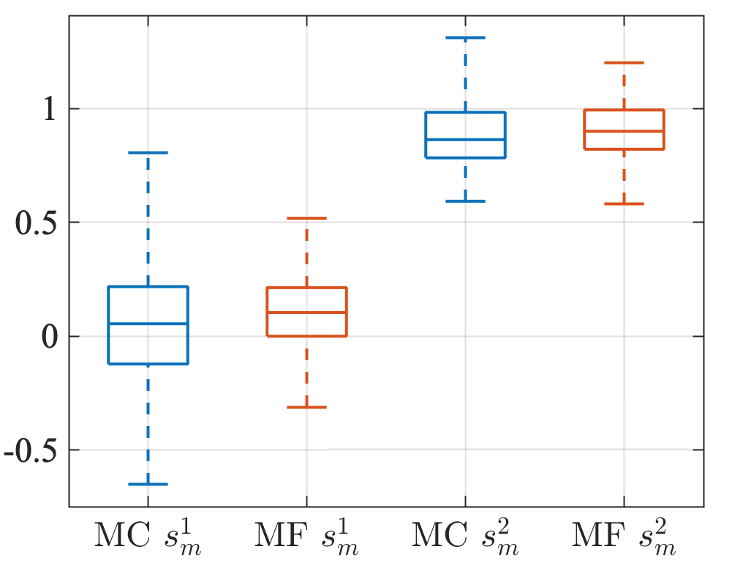}\label{sobol-a}}
   \subfloat[]{\includegraphics[width=.33\textwidth]{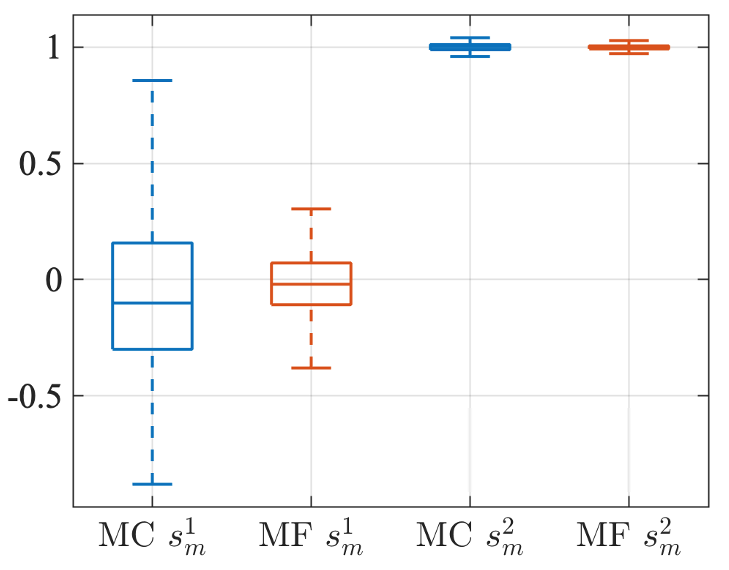}\label{sobol-b}}
    \subfloat[]{\includegraphics[width=.33\textwidth]{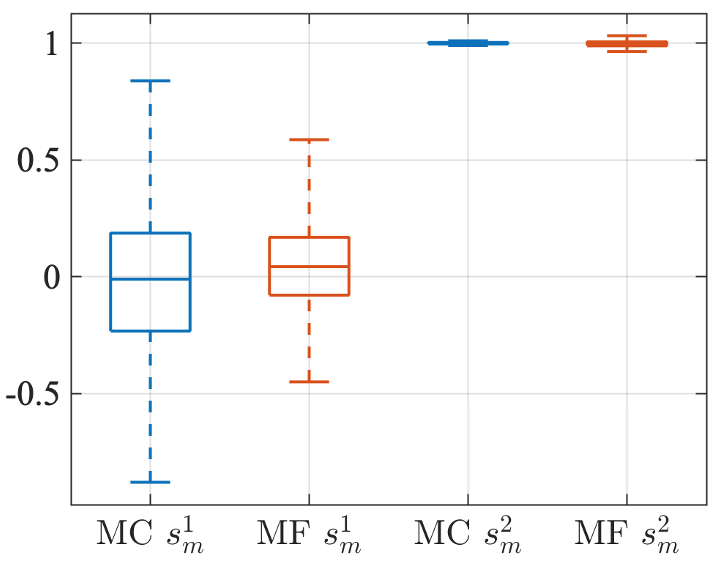}\label{sobol-c}}\quad
      \subfloat[]{\includegraphics[width=.33\textwidth]{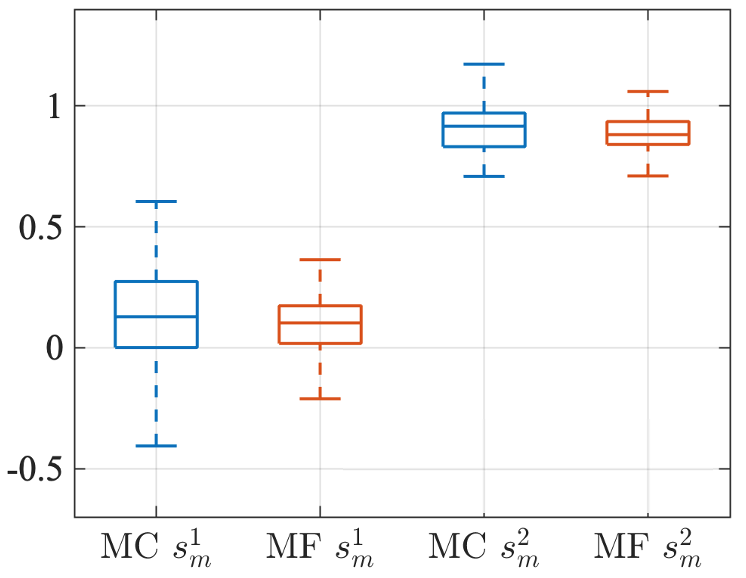}\label{sobol-d}}
   \subfloat[]{\includegraphics[width=.33\textwidth]{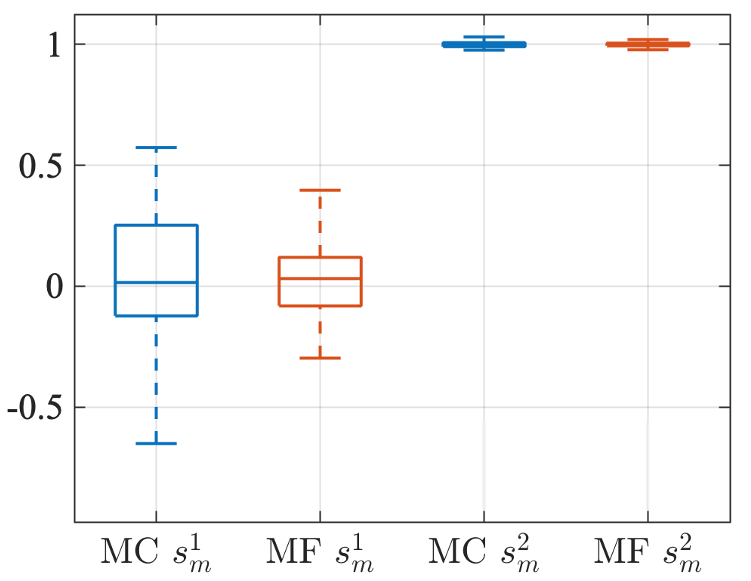}\label{sobol-e}}
    \subfloat[]{\includegraphics[width=.33\textwidth]{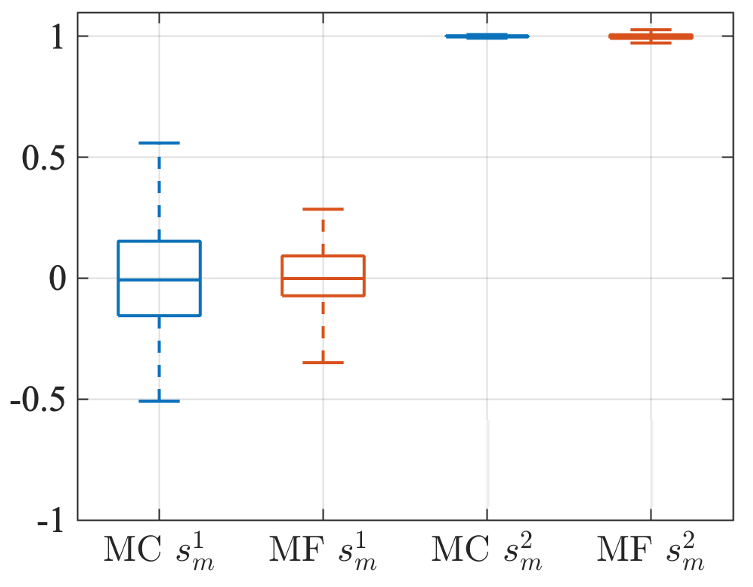}\label{sobol-f}}
  \caption[Box plots of 100 main-effect sensitivity estimate replicates for non-dimensional deposition velocity.]{Box plots of 100 main-effect sensitivity estimate replicates for non-dimensional deposition velocity with (a)-(c) $p=36$ and (d)-(f) $p=72$. $\tau_p^{+} = 0.1, 1 {\rm \ and\ }1$ from left to right. {MC using only DNS} and MFMC predictions are shown in blue and red respectively. $s_m^1$ and $s_m^2$ are main Sobol' index for $\zeta$ and $\xi$ respectively.}  
  \label{fig:sobol_verify}
\end{figure}

\subsection{Sensitivity analysis of the deposition rate}
Figure~\ref{fig:sobol_verify} presents the estimated global sensitivity of $V_{\rm dep}^{+}$ using MC with only DNS 
(high-fidelity model)
and MFMC. Data statistics of 100 estimate replicates are shown as box plots. Similar to the mean and standard deviation, the MFMC estimators admit smaller variance in the predicted Sobol' indices compared to the classic MC estimators, allowing definitive ranking of input parameters (i.e., $\zeta$ and $\xi$) with $p = 36$ while MC barely achieves this at $p = 72$. For instance, the variation in MFMC's $s_m^{1}$ with $p=36$ is smaller than MC's $s_m^{1}$ obtained with a higher budget $p=72$ across all three particle sizes.

\begin{figure}[h!]
  \centering
  \subfloat[]{\includegraphics[width=.333\textwidth]{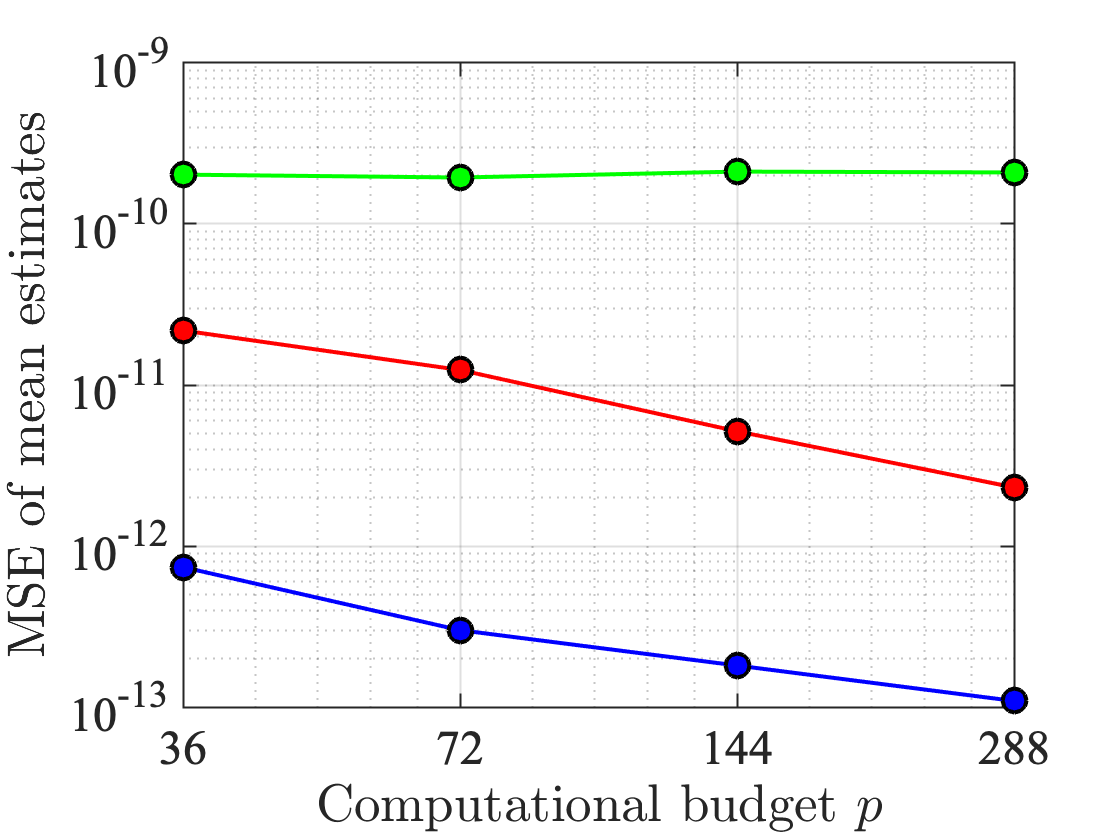}\label{mse-a}}
   \subfloat[]{\includegraphics[width=.333\textwidth]{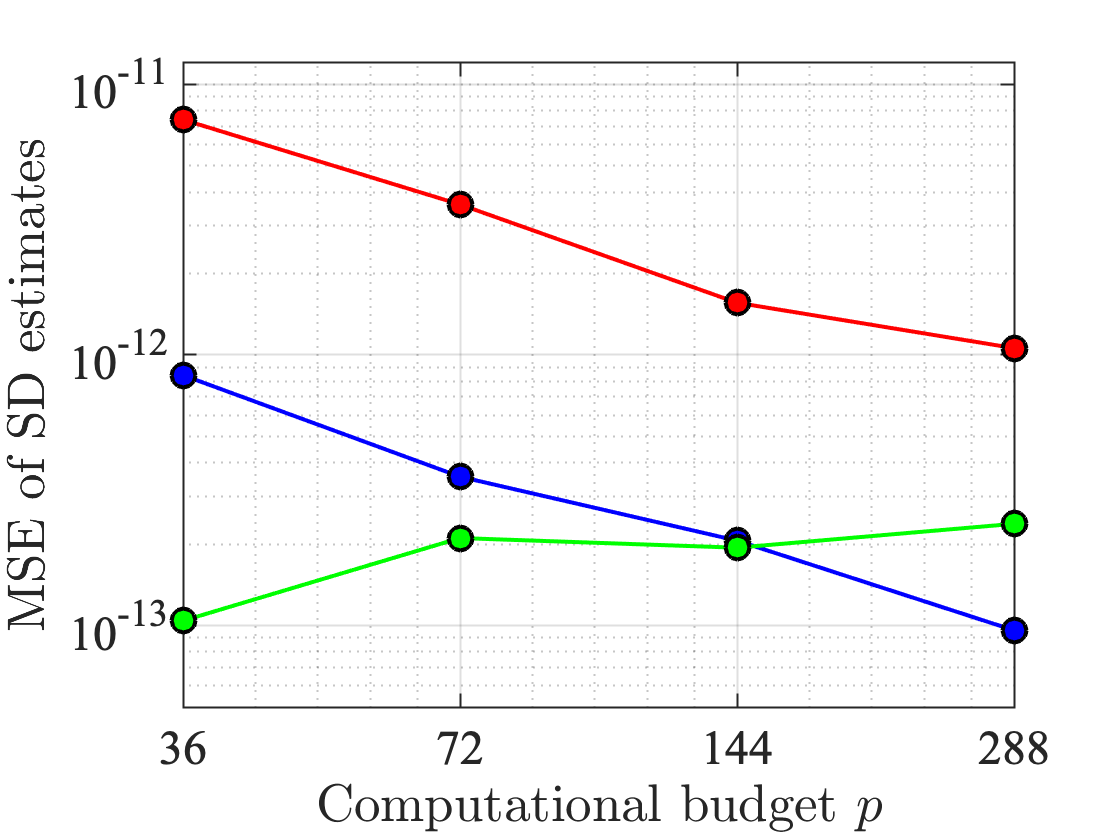}\label{mse-b}}
    \subfloat[]{\includegraphics[width=.333\textwidth]{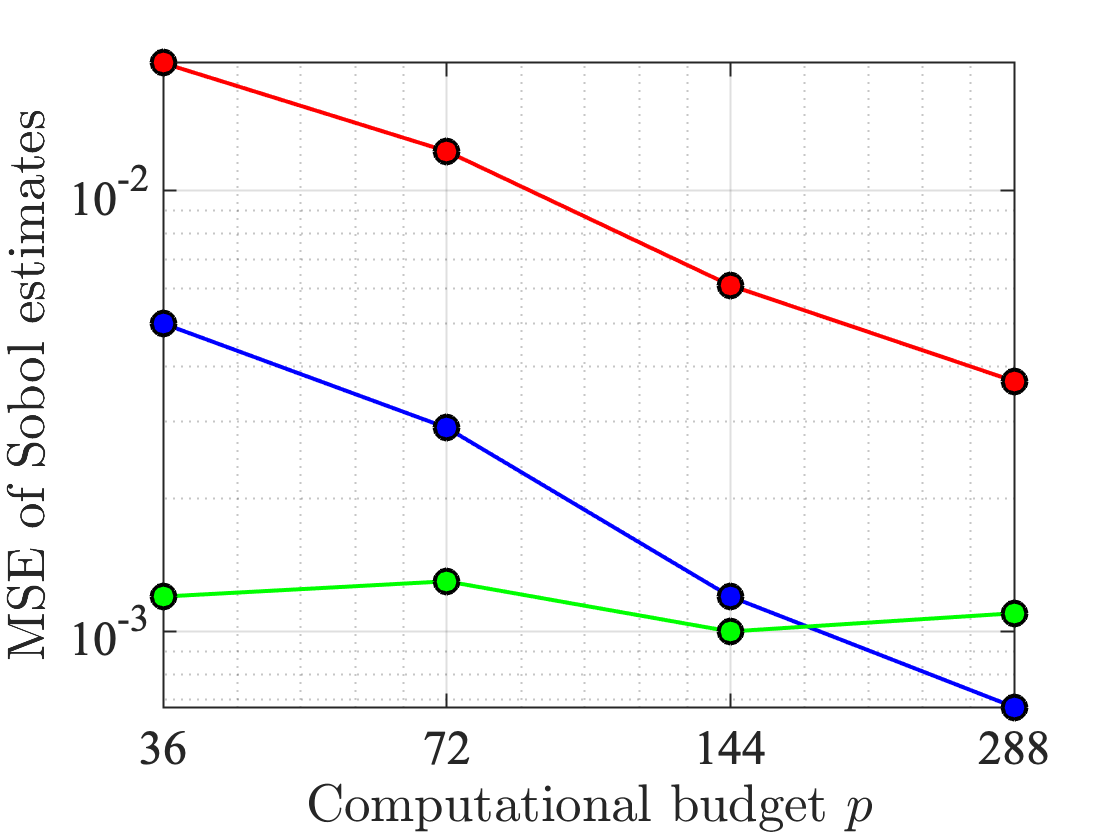}\label{mse-c}}
  \caption[The estimated MSE computed with high-fidelity only, low-fidelity only, and multi-fidelity estimators.]{The estimated MSE of MFMC estimator (blue), MC estimator with only DNS (red), and MC estimator with only one-dimensional model (green) for $\tau_p^{+} = 0.1$ and four different budgets. QoIs are (a) mean and (b) standard deviation of $V_{\rm dep}^{+}$, and (c) Sobol' index of $V_{\rm dep}^{+}$ to van der Waals.}  
  \label{fig:mse}
\end{figure}

To more quantitatively assess the benefit of  MFMC, estimated MSE is plotted in Fig.~\ref{fig:mse} for the MFMC estimator, MC estimator with only DNS, and MC estimator with only one-dimensional model. MSE is computed via Eq.~\eqref{eq:mse} with $\hat{S}$ taken from a well-converged estimates using $p=50,000$ by randomly sampling from 100 DNS data. The MFMC estimator with $p=36$ achieves the same level of MSE compared to the classic MC estimator using only DNS data with $p=288$, which translates to at least 8 times speedup. Such cost saving is seen for estimation of all three quantities: mean, standard deviation, and Sobol' index of $V_{\rm dep}^{+}$. Note that although using the low-fidelity model alone to estimate sensitivities exhibit small variation when the budget is small, it converges to a wrong solution as shown in Fig.~\ref{mse-b} and \ref{mse-c} due to the bias 
from its modeling error.
The bias in one-dimensional model prediction is also seen in Fig.~\ref{mse-a} where the error in predicted mean of $V_{\rm dep}^{+}$ is orders of magnitudes larger than MFMC.

\begin{table}
 \begin{center}
\begin{tabular}{l|l|llll}
\hline $\tau_p^{+}$ & $d_p\,(\upmu$m) & mean & standard deviation & $s_m^{1}$ & $s_m^{2}$ \\
\hline
0.1 & 1.6 & $6.66 \times 10^{-5}$ & $1.93 \times 10^{-5}\ (29.0\%)$ & 0.072 & 0.923\\
1 & 5 & $1.50 \times 10^{-3}$ & $7.65 \times 10^{-4}\ (51.0\%)$ & 0.001 & 0.998\\
10 & 16 & $5.70 \times 10^{-2}$ & $1.90 \times 10^{-2}\ (33.3 \%)$ & 0.014 & 0.990\\
\hline
\end{tabular}
\caption[Statistical estimates of deposition rates using multi-fidelity approach for three different particle sizes.]{Estimation of mean, standard deviation (as a percentage of the mean in parenthesis), main-effect Sobol' indices of $V_{\rm dep}^{+}$ using multi-fidelity approach for three different particle sizes with $p=288$.}
   \label{table:mf-final}
 \end{center}
\end{table} 

Finally, the predicted statistics of $V_{\rm dep}^{+}$ using MFMC with $p=288$ are tabulated in Table~\ref{table:mf-final}. 
Regarding uncertainty,
the deposition rate of middle-sized particles ($\tau_p^{+} =1$ or $d_p = 5\,\upmu$m) exhibits the largest standard deviation given the same variations in electrical charge and Hamaker constant, which is primarily due to these particles being most responsive to the turbulent eddies. 
Regarding sensitivity, van der Waals force has almost negligible effect on the deposition rates ($s_m^1 \approx 0$) for $\tau_p^{+}=1$ and 10, but plays a bigger role for the smallest particles ($\tau_p^{+}=1$ or $d_p = 1.6\,\upmu$m). 

\subsection{Temperature effects on particle deposition}
{The demonstration thus far 
has been focusing on
particle deposition under an all-stick assumption. }
However, both cohesion strength and temperature are known to affect the rebound mechanism of particles.
To {capture these effects,}
a physics-based model for temperature-dependent particle rebound is used~\citep{bons2017simple}, which incorporates material property sensitivity including yield stress $\sigma_y$ and normal coefficient of restitution ${\rm CoR}_n$ to temperature $T$ and normal impact velocity $V_{n1}$ observed in experiments. The model derives ${\rm CoR}_n$ based on energy conservation, which is given as
\begin{equation}
\mathrm{CoR}_{n}=\frac{V_{n 2}}{V_{n 1}}=\sqrt{\frac{\sigma_{y}^{2}}{\rho_p E}-\frac{2 A_{\mathrm{cont}} \gamma_c}{m_p}} \frac{1}{V_{n 1}}
\end{equation}
where $\sigma_y(T) = 200-0.225(T-1000)$ MPa, $\gamma_c = A/(24\pi\delta_c^2)$ is the cohesion potential energy with $\delta_c = 0.165\,$nm, $A_{\rm cont}$ is the contact area between particle and wall, $E = 70$ GPa is the Young's modulus of particle.  Similar to $\xi$ and $\zeta$, another input parameter $\theta = 0.225(T-1000)/200 \in [0,\,1]$ is introduced such that the $\sigma_y$ varies from 0 to $\sigma_{y}^{0} = 200$ MPa, the particle yield stress under typical conditions. 

\begin{figure}[h!]
  \centering
  \subfloat[]{\includegraphics[width=.45\textwidth]{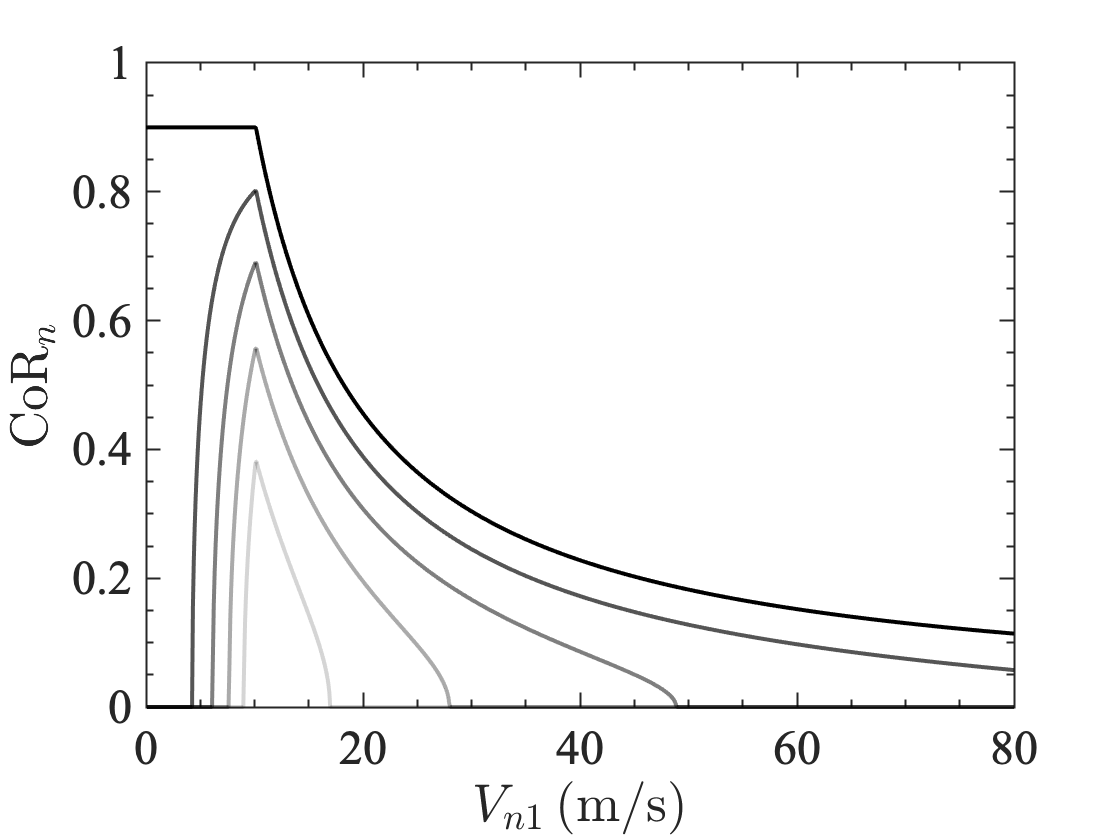}\label{fig:CoR_a}}
    \subfloat[]{\includegraphics[width=.45\textwidth]{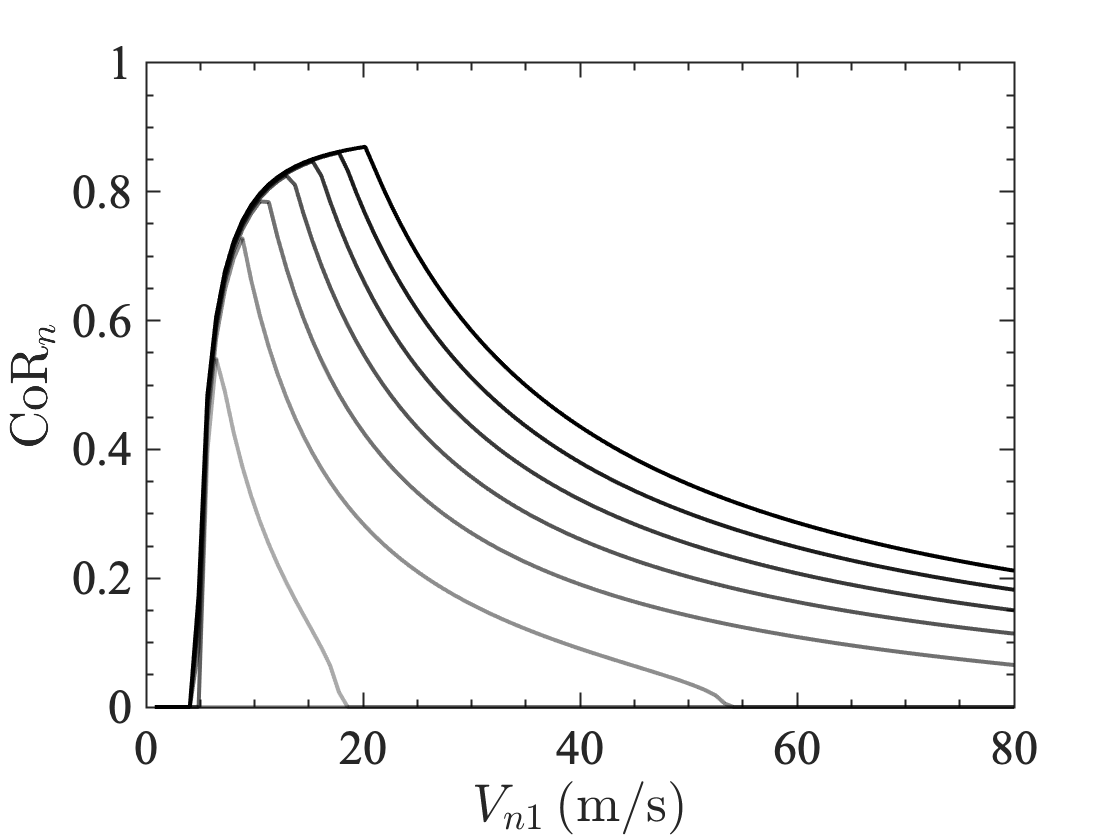}\label{fig:CoR_b}}
  \caption{Particle normal coefficient of restitution ${\rm CoR}_n$ versus normal impact velocity $V_{n1}$ predicted by the OSU rebound model~\citep{bons2017simple}. (a) $\zeta = 0$ to 1 with $\theta=0.5$ and $\tau_p^{+}=0.1$. (b) $\theta=0$ to 1 with $\zeta=1$ and $\tau_p^{+}=1$. Higher values are plotted as higher transparency.}
  \label{fig:CoR}
\end{figure}

Figure~\ref{fig:CoR} shows how ${\rm CoR}_n$ depends on $\zeta$ and $\theta$ as a function of normal impact velocity $V_{n1}$ using the rebound model of~\citet{bons2017simple}. It is observed that with increasing cohesion strength $\zeta$ or temperature $\theta$, ${\rm CoR}_n$ decreases for all impact velocities, indicative of more deposition. Here ${\rm CoR}_n=0$ represents that the particle is deposited and all-stick condition applies.

Using MFMC together with the deposition model of~\citet{bons2017simple}, we can eliminate the all-stick constraint. However, since particle rebound is not accounted in the definition of $V_{\rm dep}^{+}$, a new QoI known as capture efficiency (CE) is introduced instead. CE is defined as the rate of deposited mass $\dot{m}_{\rm dep}$ over the total mass flow rate $\dot{m}_{\rm tot}$ given by
\begin{equation}
{\rm CE} = \frac{\dot{m}_{\rm dep}}{\dot{m}_{\rm tot}} = \frac{4L}{D} \frac{u_{*}}{U_b} V_{\rm dep}^{+} \vert_{\rm stick} \le \frac{4L}{D} \frac{u_{*}}{U_b} V_{\rm dep}^{+},
\end{equation}
where $V_{\rm dep}^{+} \vert_{\rm stick}$ is the nondimensional deposition velocity of particles that do not rebound (${\rm CoR}_n=0$), the equality is achieved only when all-stick condition holds. 
CE incorporates the effect of particle rebound and is readily measurable in experiments~\citep{bons2017simple,gnanaselvam2021turbulent}.

\begin{figure}[h!]
  \centering
    \subfloat[]{\includegraphics[width=.45\textwidth]{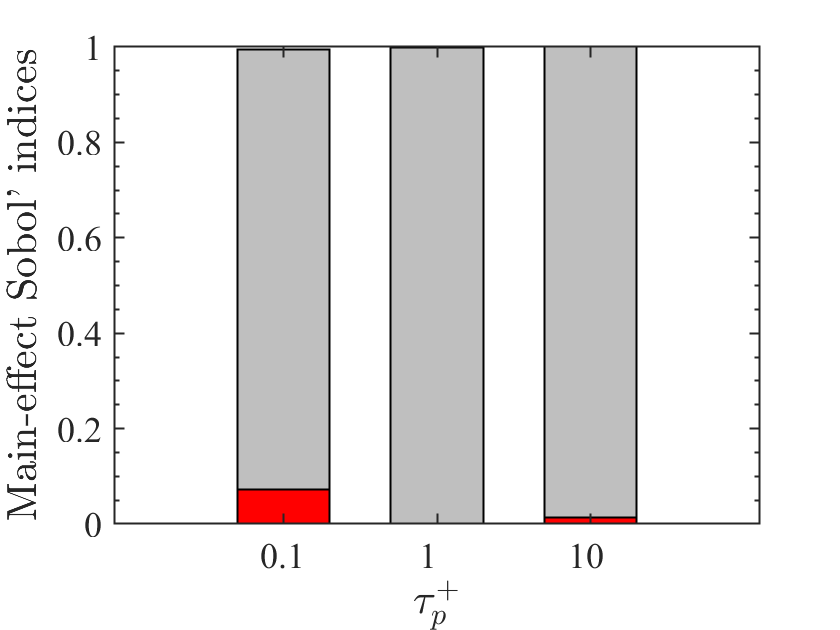}\label{fig:dep_uq_notemp}}
  \subfloat[]{\includegraphics[width=.45\textwidth]{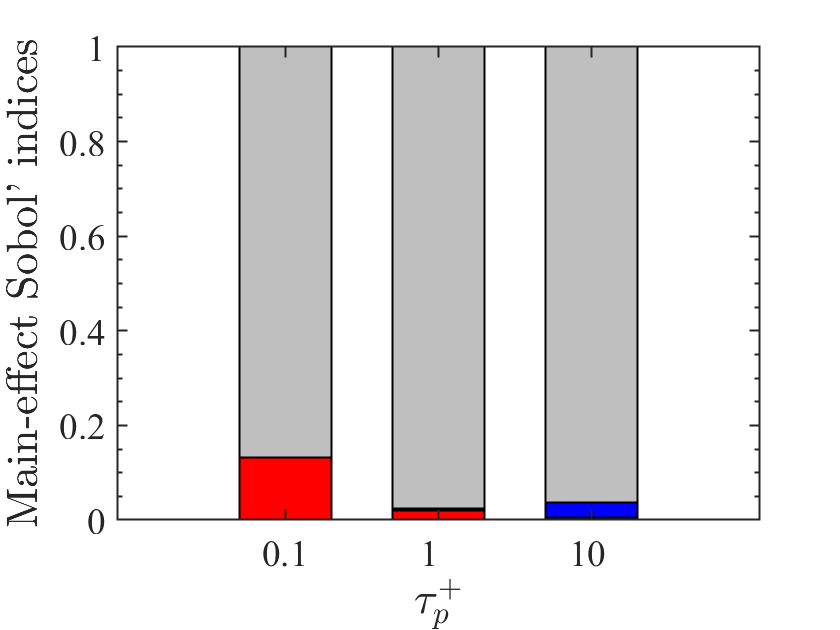}\label{fig:dep_uq_temp}}
  \caption{Main-effect Sobol' indices of particle deposition quantified by (a) $V_{\rm dep}^{+}$ and (b) CE due to variations in electrostatics, $\xi$ (gray), van der Waals, $\zeta$ (red), and temperature, $\theta$ (blue) for $\tau_p^{+}=0.1$, $1$, and $10$. }  
  \label{fig:dep_uq}
\end{figure}

The main-effect Sobol' indices of deposition, quantified by $V_{\rm dep}^{+}$ and ${\rm CE}$, due to variations in $\xi$, $\zeta$, and $\theta$ is shown in Fig.~\ref{fig:dep_uq}. It can be seen that after including the rebound physics, small particles become even more sensitive to van der Waals since adhesion reduces ${\rm CoR}_n$ and therefore prevents particles from rebound. In addition, large particles are more sensitive to temperature namely due to the decreased yield stress. This is because large particles impact the wall with higher impact velocities, which results in more plastic deformation especially at high temperatures. Nevertheless, the deposition process is still most sensitive to the variation in electrostatics. Finally, we note that although the scope of this paper is limited to particle deposition as a function of three input parameters, the same MFMC framework can be applied to study other multiphase flow problems and additional input parameters can be introduced at affordable computational costs.

\section{Conclusions}
{In this study, global sensitivity analysis was performed for particle deposition with respect to the uncertainty in electrostatic interactions, van der Waals, and particle temperature in turbulent pipe flow. In particular, Sobol' sensitivity indices, which quantify the relative impact of uncertainty from different inputs on the variability of the output QoI, were calculated.} Particle-wall contributions were found to dominate whereas particle-particle interactions due to van der Waals are negligible in dilute suspensions. Deposition predicted by the one-dimensional Eulerian model was seen to be systematically lower than DNS simulations, while both exhibit similar trends when varying particle charge and the Hamaker constant. 

A multi-fidelity Monte Carlo method was employed to optimally estimate mean, variance, and sensitivity of the deposition rate by exploiting the high correlation between DNS and the one-dimensional Eulerian model. In particular, Sobol' indices computed using MFMC offered eight times speedup compared to classic Monte Carlo. Deposition was found to be more sensitive to electrostatics than van der Waals across all particle sizes. The effect of van der Waals on deposition was seen to decay rapidly with increasing particle size. Finally by incorporating particle rebound, the effect of van der Waals was amplified for small particles and temperature variations were found to be non-negligible only for the deposition of large particles.

\section*{Acknowledgments}
This research was supported by the Office of Naval Research under Award No.~N00014-19-1-2202. 

\appendix


\section{High-fidelity model -- Direct numerical simulation}\label{sec:hf_model}

Direct numerical simulations of particle-laden pipe flows are performed in an Eulerian--Lagrangian framework. The domain is discretized on a Cartesian mesh of size $326\times256\times256$. A conservative cut-cell immersed boundary method is employed to enforce no-slip and no-penetration boundary conditions in the fluid phase at the pipe wall. Details can be found in \citet{pepiot2010direct,meyer2010conservative}. The grid spacing is chosen such that $\Delta y^{+} = \Delta z^{+} = 1.25$ and $\Delta x^{+} = 9.8$ to fully resolve the turbulence, where `$^{+}$' denotes the dimensionless wall distance defined as $\left( \cdot \right)^{+} = u_{*} \left( \cdot \right) / \nu$, where $u_{*} = \sqrt{\tau_w/\rho}$ is the friction velocity with $\tau_w$ the wall shear stress and $\rho$ the fluid density.

Particles are introduced into the flow once a statistically stationarity is reached. The maximum particle volume fraction is below $3.4\times 10^{-5}$ so that one-way coupling assumption is valid. Particle cohesion is turned on and deposition rates ($V_{\rm dep}^{+}$) are measured after the simulation is run for $240\,D/U_b$, at which the particle distribution reaches a statistically stationary state (see Fig.~\ref{fig:vis-2}). Particles are removed from the simulation after impact with the wall.

\subsection{Fluid-phase equations}
The flow of spherical particles suspended in the turbulent pipe flow is solved in an Eulerian--Lagrangian framework, where particles are treated as discrete entities of finite size and mass, and the gas phase is solved on a background Eulerian mesh. Due to the low concentrations and small particle size considered in this study, volume fraction effects and two-way coupling between the phases are neglected. The governing equations for the incompressible carrier phase are given by
\begin{equation}
\label{continuity}
\nabla\cdot\bm{u}_f=0,
\end{equation}
and
\begin{equation}
\label{velocity}
\frac{\partial\bm{u}_f}{\partial t}+ \bm{u}_f\cdot\nabla\bm{u}_f=-\frac{1}{\rho_f}\nabla p +\nu\nabla^2\bm{u}_f + \boldsymbol{F}_{\rm{IBM}}+\boldsymbol{F}_{\rm{mfr}},
\end{equation}
where  $\bm{u}_f=[u_f,~v_f,~w_f]^{\sf T}$ is the fluid velocity, $\rho_f$ is the fluid density, and $p$ and $\nu$ are the hydrodynamic pressure and kinematic viscosity, respectively. $\boldsymbol{F}_{\rm{IBM}}$ is a forcing term due to the immersed boundary to represent the pipe wall, and $\boldsymbol{F}_{\rm{mfr}}$ is uniform body force that ensures the mass flowrate remains constant~\cite{capecelatro2013eulerian}.

The equations are implemented in the framework of NGA~\citep{desjardins2008high}, a fully conservative solver tailored for turbulent flow computations. The Navier--Stokes equations are solved on a staggered grid with second-order spatial accuracy for both the convective and viscous terms, and the second-order accurate semi-implicit Crank--Nicolson scheme of~\citet{pierce2001progress} is used for time advancement.  The pressure Poisson equation that enforces continuity is solved using a multi-grid preconditioned conjugate gradient method~\citep{falgout2002hypre,van2003iterative}.

\subsection{Particle-phase equations}
The displacement of an individual particle $i$ is calculated using Newton's second law of motion
\begin{equation}
\frac{d\bm{x}_p^{(i)}}{dt} = \bm{v}_p^{(i)},\\
\end{equation}
and
\begin{equation} \label{newtonlaw}
m_p\frac{d\bm{v}_p^{(i)}}{dt}=\bm{\bm{F}}_{\text{drag}}^{(i)}+\bm{F}_{\text{col}}^{(i)}+\bm{F}_\text{vw}^{(i)}+\bm{F}_{\text{coulomb}}^{(i)},
\end{equation}
where $\bm{x}_p^{(i)}(t)$ and $\bm{v}_p^{(i)}(t)$ are the instantaneous particle position and velocity at time $t$, respectively, $m_p$ is the particle mass, $\bm{F}_{\text{drag}}^{(i)}$ is the drag force, $\bm{F}_{\text{col}}^{(i)}$ is the collision force, $\bm{F}_{\text{B}}^{(i)}$ accounts for the Brownian motion, $\bm{F}_{\text{vw}}^{(i)}$ is the van der Waals force, and  $\bm{F}_{\text{coulomb}}^{(i)}$ is the electrostatic force.

\subsubsection{Drag force}
The classic Schiller and Naumann drag correlation~\cite{clift2005bubbles} is used on the right-hand side of Eq.~\eqref{newtonlaw} to account for finite Reynolds number effects, given by
\begin{equation}\label{part_source}
\frac{\bm{\bm{F}}_{\text{drag}}^{(i)}}{m_p}=\frac{1+0.15 \text{Re}_p^{0.687}}{\tau_p}\left(\bm{u}_f[\bm{x}_p^{(i)}]-\bm{v}_p^{(i)}\right),
\end{equation}
where $\bm{u}_f[\bm{x}_p^{(i)}]$ is the fluid velocity at the location of particle $i$ and $\text{Re}_p =\| \bm{u}_f[\bm{x}_p^{(i)}]-\bm{v}_p^{(i)}\| d_p/\nu$ is the particle Reynolds number, with $d_p$ the particle diameter, and $\tau_p = \rho_pd_p^2/(18\rho_f \nu)$ is the particle response time.

\subsubsection{Collision force}\label{num_col}
Despite the low volume fractions considered here, particle collisions are needed to prevent unphysical overlap that may arise due to the attractive forces such as van der Waals force and electrostatics between oppositely charged particles. In this work, normal and tangential collisions are modeled using a modified soft-sphere approach~\cite{capecelatro2013euler} originally proposed by \citet{cundall1979discrete}. When two particles come into contact, a repulsive force is modeled as a mass-spring-damper system. In the present study, we consider inelastic collisions with a coefficient of restitution ${\rm e}{=}0.9$, representative of many solid spherical objects in dry air. To properly resolve the collisions without requiring an excessively small timestep, $\tau_{col}$ is set to be 15 times the simulation time step for all simulations presented in this work.

\subsubsection{Van der Waals force}
Cohesion of fine (sub $20\, \upmu$m) particulates strongly controls transport and deposition phenomena. Van der Waals forces are one of the dominant force contributing to surface adhesion of micron-sized particles. The van der Waals force between two particles $i$ and $j$ is modelled as $\bm{F}_{ij}^{\mathrm{vw}} = F_{ij}^{\mathrm{vw}}\,\bm{n}_{ij}$ where $F_{ij}^{\mathrm{vw}}$ is given in Eq.~\eqref{eq:vdw}. The spring stiffness $k$ used in the soft-sphere collision model described in \S~\ref{num_col} is determined based on the collision time $\tau_{\rm col}$, resulting in artificially `soft' particles to enable larger time steps. A modified van der Waals model was recently proposed to be compatible with the soft-sphere collision model~\citep{gu2016modified}.  The modification ensures the work done by the van der Waals force remains unchanged when particles overlap, such that its overall effect is insensitive to the choice of $k$ and consequently the results remain unchanged as $\Delta t$ is adjusted. This is accomplished by modifying the saturation distance and Hamaker constant when two particles are in direct contact. A value of $k_r = 7000$ N/m is used and the simulation spring stiffness $k$ is chosen such that $k_r/k=700$ as described in \citet{gu2016modified}.

It is important to note that cohesion due to van der Waals attraction and collisions are treated independently, which implicitly assumes these effects are additive in nature according to the Derjaguin, Muller and Toporov (DMT) theory~\citep{derjaguin1975effect}. The underlying assumption of the DMT theory is that cohesive forces do not modify the surface profile during particle contact. Another popular contact theory proposed by~\citet{johnson1971surface}, known as the Johnson, Kendall and Roberts (JKR) theory, assumes that adhesion occurs only within the flattened contact region such that the collision force is nonlinearly coupled with the cohesion force and consequently cannot be treated as additive. As suggested by~\citet{johnson1997adhesion}, the DMT approximation is valid when $\lambda_T \ll 1$ and the JKR model is valid when $\lambda_T \gg 1$, with $\lambda_T$ the dimensionless Tabor parameter defined as
\begin{equation}\label{eq:tabor}
\lambda_T = \left( \frac{2d_p\gamma_c^2}{E^2\delta_c^3} \right)^{1/3}.
\end{equation}
In this work, particle contact mechanics are based on the DMT theory due to the low values of $\lambda_T$ under consideration and to be consistent with the cohesion model of \citet{gu2016modified}.

\subsubsection{Electrostatic force}
The last term in Eq.~\eqref{newtonlaw} is the electrostatic force governed by Coulomb's law according to $\bm{F}_{\mathrm{coulomb}}^{(i)} = \sum_{j\ne i} F_{ij}^{\mathrm{C}}\,\bm{n}_{ij}$, where $F_{ij}^{\mathrm{C}}$ is given by Eq.~\eqref{eq:coulomb}. For the simulations considered in this work, the electrical permittivity is assumed constant and taken to be $\epsilon_0$. The force of interaction between the particles is attractive if their charges have opposite signs and repulsive if like-signed. As shown in Eq.~\eqref{eq:coulomb}, a direct summation will result in $\mathcal{O}(N^2)$ computations.

To avoid the $\mathcal{O}\left(N^2\right)$ calculation, the electrostatic force is computed according to 
\begin{equation}
\bm{F}_{\text{coulomb}}^{(i)} = q_p^{(i)} \bm{E}[\bm{x}_p^{(i)}],
\label{coulomb_force}
\end{equation}
where $\bm{E}[\bm{x}_p^{(i)}] $ is the electric field interpolated to the position of particle $i$. The electric field is solved using the particle-particle particle-mesh (P$^3$M) method introduced by \citet{hockney}, which separate long-range and short-range contributions by a cutoff radius $r_{\max}$. The long-range field is solved using Fast Fourier Transform (FFT)~\citep{press1992numerical} on an underlying mesh. With a well-chosen cutoff radius, it can be shown that the overall computational cost is $\mathcal{O}\left(N \log{N}\right)$. This method has been demonstrated to provide accurate results when considering attractive inter-particle forces due to electrostatics~\citep{yao2018competition} and we recently extended it to non-periodic geometries while retaining its high accuracy and cost savings~\citep{yao2021accurate}.

\section{Low-fidelity model -- one-dimensional Eulerian deposition model}\label{sec:lf_model}

\subsection{A unified Eulerian formulation of turbulent deposition}

\citet{guha1997unified} established, by deriving from the fundamental Eulerian conservation equations of mass and momentum for the particles, a unified advection-diffusion theory in which turbophoresis arises naturally. The theory includes molecular and turbulent diffusion, thermophoresis, shear-induced lift force, electrical forces, and gravity. The predicted deposition rates by this one-dimensional model have been shown to be comparable to Lagrangian calculations~\citep{yao2021accurate} and is therefore ideal for multi-fidelity uncertainty quantification. Here we present a model formulation based on~\citet{guha1997unified} while incorporating boundary conditions derived from kinetic theory~\cite{young1997theory}. The model is also modified to include additional source terms due to cohesion in the momentum equation.

Most of experimental and numerical study on pipe deposition report non-dimensionalized deposition velocity versus particle relaxation time defined as
\begin{equation}
V_{dep}^{+} \equiv \frac{V_{dep}}{u_{\tau}} = \frac{J_{w}}{\rho_{p m} u_{\tau}}
\end{equation}
\begin{equation}
\tau_p^{+} \equiv \frac{\tau_p u_{\tau}^{2}}{\nu}=\frac{2}{9}\left(\frac{\rho_p^{0}}{\rho_f}\right) \frac{r^{2} u_{\tau}^{2}}{\nu^{2}}
\end{equation}
where $J_w$ is the particle mass flux to the wall per unit area, $\rho_{pm}$ is the mean particle density in the pipe and $u_\tau = \sqrt{\tau_w/\rho_g}$ is the friction velocity where $\tau_w$ is the wall shear stress, $\rho_f$ is the gas density, $\rho_p^0$ is the  particle material density.

The non-dimensionalized governing equations are given by
\begin{equation}\label{eq:momentum}
\overline{V}_{{py}}^{+} \frac{\partial \overline{V}_{{py}}^{+}}{\partial y^{+}}+\frac{\overline{V}_{{py}}^{+}}{\tau_{1}^{+}}=-\frac{\partial}{\partial y^{+}}(\mathfrak{R} \overline{V_{{f} y}^{\prime+2}})
\end{equation} 
\begin{equation}\label{eq:continuity}
\partial V_{{dep}}^{+} / \partial y^{+} = 0, \quad V_{{dep}}^{+}=-\left(\frac{D_{\mathrm{B}}}{\nu}+\frac{\varepsilon}{\nu}\right) \frac{\partial \rho_{{p}}^{+}}{\partial y^{+}}-\rho_{{p}}^{+} D_{\mathrm{T}}^{+} \frac{\partial \ln T}{\partial y^{+}}+\rho_{{p}}^{+} \overline{V}_{{py}}^{+}
\end{equation}
where $\overline{V}_{{py}}^{+} =\overline{V}_{{py}} / u_{\tau}$, $\rho_{{p}}^{+} =\overline{\rho}_{{p}} / \rho_{{p}_{0}}$, $V_{{f} y}^{\prime+} =V_{{f} y}^{\prime} / u_{\tau}$, $D_{T}^{+} =D_{T} / \nu$. Here $y^{+} = y\,u_\tau / \nu$ is the dimensionless wall unit and $ \rho_{{p}_{0}}$ is the particle density at the pipe centerline. Equations are solved numerically with grid refinement close to wall, the drag term is treated implicitly to assure numerical stability. The boundary condition is derived from kinetic theory by~\citet{young1997theory}
\begin{equation}
-\left(\frac{D_{\mathrm{B}}}{\nu}+\frac{\varepsilon}{\nu}\right)  \frac{\partial \rho_{{p}_w}^{+}}{\partial y^{+}} = \rho_{{p}_w}^{+}\overline{V}_{py}^{+} \left( \frac{\exp{(-{M_r^2})}}{2\sqrt{\pi} {M_r}} - \frac{1-\text{erf}({M_r})}{2} \right)
\end{equation}
where $\rho_{{p}_w}$ is the particle density at the pipe wall. Details on the empirical closures for $\mathfrak{R}$, $V_{{f} y}^{\prime}$, $D_B$,  $D_T$, $\varepsilon$ are described in~\citet{guha1997unified,guha2008transport}.

\subsection{Modeling of particle-wall interactions}
In the presence of van der Waals and electrostatic forces, the Reynolds-averaged particle momentum equations (Eq.~\eqref{eq:momentum}) is modified to include these two additional source terms given as
\begin{equation}\label{eq:mom_modified}
\overline{V}_{{py}}^{+} \frac{\partial \overline{V}_{{py}}^{+}}{\partial y^{+}}+\frac{\overline{V}_{{py}}^{+}}{\tau_{1}^{+}}=-\frac{\partial}{\partial y^{+}}(\mathfrak{R} \overline{V_{{f} y}^{\prime+2}}) + F_{\rm elec}^{+} + F_{\rm vdw}^{+},
\end{equation}
where the electrostatic force and van der Waals force are non-dimensionalized as $F_{\rm elec}^{+} = F_{\rm elec} \,\nu / u_\tau^3$ and $F_{\rm vdw}^{+} = F_{\rm vdw} \,\nu / u_\tau^3$ respectively. The charged particle is assumed to experience an attractive force due to induced charges on the wall, which is modeled as image charging given as
\begin{equation}
F_{\rm elec}=-\frac{3q_p^{2}}{8 \pi^{2} \varepsilon_{0} \rho_{\mathrm{p}}^{0}\,d_p^{3}\,y^{2}}.
\end{equation}
The van der Waals force between a spherical particle and wall is given as
\begin{equation}
F_{\rm vdw}=-\frac{A \,d_p}{12 m_p\,(y-0.5d_p)^2}.
\end{equation}
Note that when particles are in contact with the wall, $y = 0.5d_p$ and $F_{\rm vdw}$ becomes infinity. To avoid this singularity, a cutoff distance $10^{-12}\,$m is used when solving Eq.~\eqref{eq:mom_modified} numerically, which was found to be the maximum value that was insensitive to the predicted deposition rates.

\bibliographystyle{elsarticle-num-names} 

\end{document}